\documentclass[amsmath, aps,prc,twocolumn,groupedaddress, showkeys, showpacs]{revtex4}
\usepackage{graphicx,epsfig, longtable}

\begin{document}

\title{Unified description of neutron superfluidity in the neutron-star crust with 
analogy to anisotropic multi-band BCS superconductors}

\author{N.~Chamel}
\author{S.~Goriely}
\affiliation{Institut d'Astronomie et d'Astrophysique, CP-226, Universit\'e Libre de Bruxelles, 
1050 Brussels, Belgium}
\author{J.M.~Pearson}
\author{M.~Onsi}
\affiliation{D\'ept. de Physique, Universit\'e de Montr\'eal, Montr\'eal (Qu\'ebec), H3C 3J7 Canada}
\date{\today}

\begin{abstract}
The neutron superfluidity in the inner crust of a neutron star has been 
traditionally studied considering either homogeneous neutron matter 
or only a small number of nucleons confined inside the spherical 
Wigner-Seitz cell. Drawing analogies with the recently discovered 
multi-band superconductors, we have solved the anisotropic multi-band 
BCS gap equations with Bloch boundary conditions, thus providing a 
unified description taking consistently into account both the free neutrons 
and the nuclear clusters. Calculations have been carried out using the 
effective interaction underlying our recent Hartree-Fock-Bogoliubov nuclear 
mass model HFB-16. We have found that even though the presence of inhomogeneities 
lowers the neutron pairing gaps, the reduction is much less than that predicted by 
previous calculations using the Wigner-Seitz approximation. 
We have studied the disappearance of superfluidity with increasing temperature.  
As an application we have calculated the neutron specific heat, which is an 
important ingredient for modeling the thermal evolution of newly-born neutron 
stars. This work provides a new scheme for realistic calculations of superfluidity 
in neutron-star crusts.
\end{abstract}


\maketitle

\section{Introduction}

The possibility of superfluidity inside neutron stars was suggested a long time ago by 
Migdal~\cite{mig59}, only two years after the formulation of the theory of electron 
superconductivity by Bardeen, Cooper and Schrieffer (BCS)~\cite{bcs57} and before the 
discovery of the first pulsars. This prediction was later supported by the observation 
of the long relaxation time of the order of months following the first observed glitch 
in the Vela pulsar~\cite{baym69}. Glitches were subsequently observed in other pulsars. 
Pulsar glitches are believed to be related to the dynamics of the neutron superfluid permeating 
the inner layers of the solid neutron star crust~\cite{and75, chac06, kos09}. Understanding the 
properties of this neutron superfluid is also of prime importance for modeling the cooling 
of newly-born neutron stars~\cite{lat94, gne01,monr07,for09} and strongly magnetised neutron stars~\cite{agui09}, 
the thermal relaxation of quasi-persistent X-ray transients~\cite{shter07, brown09} or 
the quasi-periodic oscillations recently detected in the giant flares of Soft-Gamma 
Repeaters~\cite{car05,cha05,lars09}.

So far most microscopic studies of neutron superfluidity have been devoted to the case of 
uniform infinite neutron matter~\cite{dean03}. However the coherence length of the neutron superfluid in the 
neutron-star crust is typically smaller than the lattice spacing and may even be comparable 
to the size of the nuclear clusters in some layers~\cite{blas97, mat06} (see also Section 8.2.3 
of Ref.~\cite{lrr}). This situation is in sharp contrast to that encountered in ordinary type I 
electron superconductors, for which the electron Cooper pairs are spatially extended over 
macroscopic distances so that the order parameter is essentially uniform~\cite{tin96} (note that neutron stars are 
much too hot for electrons to be superconducting there, see for instance the discussion in Section 
8.1 of Ref.~\cite{lrr}).
The effects of the inhomogeneities on neutron superfluidity in neutron-star crust have been 
studied in the mean-field approximation with realistic nucleon-nucleon potentials~\cite{bar97} and 
effective nucleon-nucleon interactions~\cite{bar98, mont04,sand04, khan05,monr07, for09}. 
Systematic fully self-consistent calculations in the entire inner crust (at zero temperature) 
have been recently carried out using semi-microscopic energy functionals~\cite{bal07}.
In this latter work, it was found that even though inhomogeneities are small in the 
densest regions of the crust, the $^1$S$_0$ neutron pairing gaps are strongly reduced compared 
to those in uniform neutron matter. Moreover the weaker the pairing force is, the 
stronger is the suppression of the gaps. In somes cases, the gaps even almost completely vanish. 
All these quantum calculations have been carried out in the Wigner-Seitz (W-S) approximation 
according to which the lattice is decomposed into a set of identical spherical cells 
centered around each cluster. The radius of each sphere is chosen so that its volume is 
equal to $1/\rho_{\rm N}$, where $\rho_{\rm N}$ is the cluster density (number of lattice sites
per unit volume). As discussed in Ref.~\cite{bal06}, two different boundary conditions 
yielding an almost constant neutron density $\rho_n(r)$ near the cell edge, can be chosen. 
While the difference in the predicted pairing gaps are small for average 
nucleon densities $\bar \rho\lesssim 0.03$ fm$^{-3}$, the uncertainties become increasingly 
large in the deeper layers of the crust~\cite{bal07}. This limitation of the W-S method is 
related to the existence of spurious neutron shell effects due to the discretisation 
of the single-particle (s.p.) energy spectrum~\cite{cha07,bal06}. 

In this paper, we present the first calculations of neutron superfluidity in neutron-star 
crusts going beyond the W-S approach by using the BCS theory of anisotropic multi-band 
superconductivity. This theory is briefly reviewed in Section~\ref{sec1}. Our model 
of the neutron-star crust are discussed in Section~\ref{sec2}. The solutions of the BCS equations are
presented in Section~\ref{sec4}. We have focused on the deepest regions of the crust, 
for average nucleon densities $\bar \rho$ between $0.05$ and $0.07$ fm$^{-3}$, where 
the results from the W-S approximation are the most uncertain~\cite{bal07}. 
In Section~\ref{sec5}, we discuss the validity of the local density approximation (LDA) and the importance 
of proximity effects. The disappearance of superfluidity with increasing temperature
is studied in Sections~\ref{sec6} and \ref{sec7}. In Section~\ref{sec8}, we present 
numerical results of the neutron specific heat.

\section{BCS theory of superfluid neutrons in neutron-star crusts} 
\label{sec1}
The standard formulation of the BCS theory starts with the Hamiltonian~\cite{bcs57} 
\begin{multline}
\label{eq.1}
H=\sum_{\sigma, \alpha, \pmb{k}}  (\varepsilon_{\alpha\pmb{k}}-\mu)\, c_{\alpha\pmb{k}\sigma}^\dagger c_{\alpha\pmb{k}\sigma}\\
+\sum_{\alpha, \beta, \pmb{k},\pmb{k^\prime}} V_{\alpha\pmb{k}\beta\pmb{k^\prime}}\,  c^\dagger_{\alpha\pmb{k}\uparrow} c^\dagger_{\alpha -\pmb{k}\downarrow} c_{\beta-\pmb{k^\prime}\downarrow} c_{\beta \pmb{k^\prime}\uparrow}
\end{multline}
where $c^\dagger_{\alpha \pmb{k}\sigma}$ ($c_{\alpha \pmb{k}\sigma}$) are the creation 
(annihilation) operators for Bloch states with wave vector $\pmb{k}$, band index $\alpha$ 
and spin $\sigma$, $\varepsilon_{\alpha\pmb{k}}$ are the s.p. energies (assumed to be independent
of the spin state), $\mu$ the chemical potential and $V_{\alpha\pmb{k}\beta\pmb{k^\prime}}$ 
are the matrix elements of the two-body pairing interaction.
In the mean-field approximation at finite temperature, the quasi-particle (q.p.) energies are 
given by
\begin{equation}
\label{eq.2}
E_{\alpha\pmb{k}}=\sqrt{(\varepsilon_{\alpha\pmb{k}}-\mu)^2 +\Delta_{\alpha\pmb{k}}^2}
\end{equation}
where 
$\Delta_{\alpha\pmb{k}}$ are solutions of the anisotropic multi-band BCS 
gap equations (setting the Boltzmann constant  $k_{\rm B}=1$) 
\begin{equation}
\label{eq.3}
\Delta_{\alpha\pmb{k}} = - \frac{1}{2} \sum_{\beta} \sum_{\pmb{k^\prime}} V_{\alpha \pmb{k}\beta \pmb{k^\prime}}
 \frac{\Delta_{\beta\pmb{k^\prime}}}{E_{\beta \pmb{k^\prime}} } \tanh \frac{E_{\beta \pmb{k^\prime}}}{2T}\, .
\end{equation}

In conventional superconductors the pairing interaction is induced by electron-phonon
coupling~\cite{tin96}. It is usually a very good approximation to take the matrix elements 
$V_{\alpha \pmb{k}\beta \pmb{k^\prime}}$ as constant and non-zero only 
within a small energy shell of the order $\sim \hbar \omega_{\rm p}$ around the Fermi level, 
where $\omega_{\rm p}$ is the ion-plasma frequency. In this case the gap parameters 
$\Delta_{\alpha\pmb{k}}$ depend neither on the band index $\alpha$ nor on the wave vector 
$\pmb{k}$ and are all equal to a single constant, the pairing gap $\Delta$~\cite{bcs57}. The 
possibility of multi-band superconductors characterized by the existence of several pairing 
gaps $\Delta_{\alpha}$ was raised soon after the formulation of the BCS theory~\cite{suhl59}, 
but clear experimental evidence was lacking until the discovery in 2001 of superconductivity in 
magnesium diboride, whose unusual properties can be nicely explained by a two-band model~\cite{choi02,iav02}. 
Since then many other multi-band superconductors have been found such as the iron pnictide 
superconductors~\cite{hun08}. In these materials, several bands can intersect the Fermi level, yielding 
a complex multi-sheeted Fermi surface. Pairing is still thought to be mediated by the 
exchange of phonons but the electrons on the different sheets of the Fermi surface feel very different 
pairing interactions leading to the existence of different gaps.
In neutron-star crust, the formation of neutron pairs giving rise to superfluidity is directly 
triggered by the strong neutron-neutron interaction which is always attractive at low densities in 
the $^1$S$_0$ channel. The number of bands contributing appreciably to the pairing gap $\Delta_{\alpha\pmb{k}}$ 
can thus be huge (about $\sim 10^2-10^3$ in the dense layers of the inner crust considered in this work). 
Since the matrix elements of the pairing force may \textit{a priori} vary appreciably, we have solved Eqs.~(\ref{eq.3}) 
in the most general case.

\section{Model of neutron-star crust}
\label{sec2}

We have determined the equilibrium structure and composition of the inner crust of neutron stars
by using the fourth-order Extended Thomas-Fermi method with quantum shell effects added via the 
Strutinsky-Integral theorem. This so-called ETFSI method, as applied to the equation of state of 
neutron-star crusts, has been described in detail in Ref.~\cite{onsi08}. It is a high-speed approximation 
to the self-consistent Hartree-Fock method. We have neglected the small neutron shell effects~\cite{oy94}, 
the estimation of which in current calculations is plagued by the approximate treatment of the interaction 
between the unbound neutrons and the nuclear lattice~\cite{cha07,bal06}. 
The calculations have been carried out using an effective nucleon-nucleon interaction of the Skyrme type
\begin{eqnarray}
\label{eq.4}
v^{\rm Sky}(\pmb{r_i}, \pmb{r_j}) & = & t_0(1+x_0 P_\sigma)\delta({\pmb{r}_{ij}}) \nonumber \\
&+&\frac{1}{2} t_1(1+x_1 P_\sigma)\frac{1}{\hbar^2}\left[p_{ij}^2\,\delta({\pmb{r}_{ij}})
+\delta({\pmb{r}_{ij}})\, p_{ij}^2 \right]\nonumber\\
&+&t_2(1+x_2 P_\sigma)\frac{1}{\hbar^2}\pmb{p}_{ij}.\delta(\pmb{r}_{ij})\,
 \pmb{p}_{ij} \nonumber\\
&+&\frac{1}{6}t_3(1+x_3 P_\sigma)\rho(\pmb{r})^\gamma\,\delta(\pmb{r}_{ij})
\nonumber\\
&&\frac{\rm i}{\hbar^2}W_0(\mbox{\boldmath$\sigma_i+\sigma_j$})\cdot
\pmb{p}_{ij}\times\delta(\pmb{r}_{ij})\,\pmb{p}_{ij}  \quad ,
\end{eqnarray}
where $\pmb{r}_{ij} = \pmb{r}_i - \pmb{r}_j$, $\pmb{r} = (\pmb{r}_i + 
\pmb{r}_j)/2$, $\pmb{p}_{ij} = - {\rm i}\hbar(\pmb{\nabla}_i-\pmb{\nabla}_j)/2$
is the relative momentum, and $P_\sigma$ is the two-body 
spin-exchange operator and $\rho(\pmb{r})$ is the total nucleon density at position $\pmb{r}$. 
The pairing interaction that we take here acts only between nucleons of the same charge state 
$q$ ($q = n$ or $p$ for neutron or proton, respectively) 
and is given by
\begin{equation}
\label{eq.9}
v^{\rm pair}_q(\pmb{r_i}, \pmb{r_j})= 
v^{\pi\, q}[\rho_n(\pmb{r}),\rho_p(\pmb{r})]~\delta(\pmb{r_{ij}})\, ,
\end{equation}
where $\pmb{r_{ij}} = \pmb{r_i} - \pmb{r_j}$, $\pmb{r} = (\pmb{r_i} + 
\pmb{r_j})/2$ and $\rho_n(\pmb{r})$ and $\rho_p(\pmb{r})$ are the neutron and 
proton density at position $\pmb{r}$ respectively.

\begin{figure}
\includegraphics[scale=0.3]{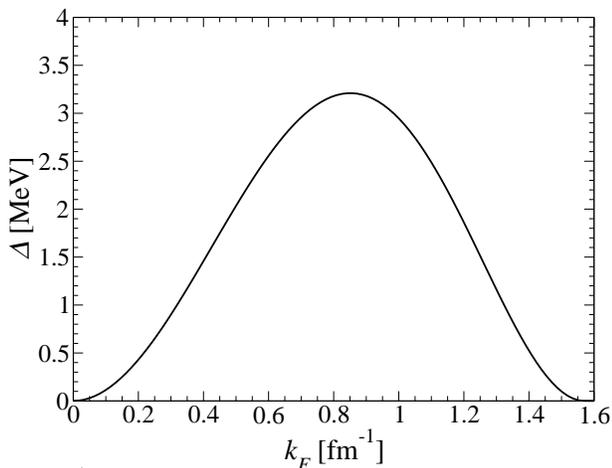}
\vskip -0.5cm
\caption{$^1$S$_0$ neutron pairing gap in infinite uniform neutron matter vs 
Fermi wave number $k_{\rm F}=(3\pi^2 \rho_n)^{1/3}$, as used in model HFB-16~\cite{cha08}.}
\label{fig1}
\end{figure}

We have adopted the parametrisation BSk16, underlying the HFB-16 nuclear mass model~\cite{cha08}. 
The parameters of this force are given in Table~\ref{tab1}. This force is particularly suitable 
for studying neutron-rich environments such as neutron-star crusts since it has been constrained 
to reproduce the equation of state and the $^1S_0$ pairing gap of infinite homogeneous neutron matter, 
as calculated for the realistic Argonne v$_{14}$ potential and shown in Fig.~\ref{fig1}. The expression 
of $v^{\pi\, q}[\rho_n(\pmb{r}),\rho_p(\pmb{r})]$ can be found in Ref.~\cite{cha08}. 
Moreover, we can hope that the nuclear inhomogeneities in the neutron-star crust will be properly taken 
into account, given the excellent fit to essentially all the available experimental nuclear mass data.

In order to solve the BCS Eqs.~(\ref{eq.3}), we first need to determine the neutron s.p. energies 
$\varepsilon_{\alpha\pmb{k}}$ and wavefunctions $\varphi_{\alpha\pmb{k}}(\pmb{r})$. For this purpose, 
we have solved the following three-dimensional Schroedinger equation
\begin{equation}
\label{eq.5}
-\pmb{\nabla}\cdot \frac{\hbar^2}{2 M_n^*(\pmb{r})} 
\pmb{\nabla}\varphi_{\alpha\pmb{k}}(\pmb{r}) + U_n(\pmb{r})\varphi_{\alpha\pmb{k}}(\pmb{r}) 
= \varepsilon_{\alpha\pmb{k}} \varphi_{\alpha\pmb{k}}(\pmb{r})\, .
\end{equation}
The effective mass $M_n^*(\pmb{r})$ and the potential $U_n(\pmb{r})$ are given by
(we neglect the small rearrangement term coming from the pairing force, see for instance~\cite{cha08})
\begin{widetext}
\begin{eqnarray}
\label{eq.7}
\frac{\hbar^2}{2 M_n^*} &=& \frac{\hbar^2}{2 M_n} + \frac{1}{4} t_1 \Biggl[ \left(1+ \frac{1}{2} x_1\right) \rho 
- \left( \frac{1}{2} + x_1\right)\rho_n\Biggr] + \frac{1}{4} t_2 \Biggl[ \left(1+ \frac{1}{2} x_2\right) \rho 
+ \left( \frac{1}{2} + x_2\right)\rho_n\Biggr] \, ,
\end{eqnarray}
\begin{eqnarray}
\label{eq.8}
U_n&=& t_0\Biggl[\left(1+ \frac{1}{2} x_0\right)\rho - \left(\frac{1}{2} +x_0\right)\rho_n\Biggr]+\frac{1}{4} t_1 \Biggl[\left(1+ \frac{1}{2} x_1\right) \left(\tau - \frac{3}{2}\nabla^2\rho\right) - \left(\frac{1}{2} +x_1\right)\left(\tau_n - \frac{3}{2}\nabla^2\rho_n\right) \Biggr]  \nonumber \\
&+&\frac{1}{4} t_2 \Biggl[\left(1+ \frac{1}{2} x_2\right) \left(\tau + \frac{1}{2}\nabla^2\rho\right) + \left(\frac{1}{2} +x_2\right)\left(\tau_n + \frac{1}{2}\nabla^2\rho_n\right) \Biggr]\nonumber\\
&+&\frac{1}{12} t_3 \Biggl[\left(1+ \frac{1}{2} x_3\right) (2+\gamma)\rho^{\gamma+1} - \left(\frac{1}{2} +x_3\right)\biggl(2\rho^\gamma \rho_n+\gamma \rho^{\gamma-1} (\rho_n^2+\rho_p^2) \biggr) \Biggr]\, ,
\end{eqnarray}
\end{widetext}
where $\tau_n(\pmb{r})$ and $\tau_p(\pmb{r})$ are the neutron and proton 
kinetic-energy densities respectively (we have introduced $\tau=\tau_n+\tau_p$). 
Both $M_n^*(\pmb{r})$ and $U_n(\pmb{r})$ were determined using the ETFSI fields. 
In Eq.~(\ref{eq.5}), we have neglected the spin-orbit coupling arising from the 
last term of Eq.(\ref{eq.4}). This approximation is justified because the spin-orbit
coupling is proportional to $\nabla \rho_n$ and $\nabla \rho_p$ (see for instance 
Appendix A of Ref.~\cite{cha08}). In neutron-star crust, nuclear clusters have a 
very diffuse surface as can be seen in Fig.\ref{fig2}, and consequently the spin-orbit 
coupling is much smaller than that in isolated nuclei~\cite{cha07}. 

Although neutron shell effects represent a small correction to the total energy density~\cite{oy94}, they are 
expected to have a much stronger impact on neutron superfluidity due to the highly non-linear 
nature of the pairing phenomenon. This is the reason why we have not followed the usual practice of 
applying the W-S approximation for solving Eq.~(\ref{eq.5}), but have imposed the Bloch boundary conditions
\begin{equation}
\label{eq.6}
\varphi_{\alpha\pmb{k}}(\pmb{r}+\pmb{\ell})=\exp({\rm i}\, \pmb{k}\cdot\pmb{\ell})\varphi_{\alpha\pmb{k}}(\pmb{r})  \, , 
\end{equation}
where $\pmb{\ell}$ denotes any lattice translation vector. Note that in this case, Eq.~(\ref{eq.5}) 
has to be solved for each wave vector $\pmb{k}$ while in the W-S method only a single wave vector 
is considered, namely $\pmb{k}=0$ (see Ref.~\cite{cha07} for a discussion about the W-S approximation). 
Following the standard assumptions, we have considered a body-centered cubic lattice~\cite{lrr}. The BCS 
Hamiltonian~(\ref{eq.1}) can then be obtained from the s.p. states, once the pairing interaction has been specified. 
The matrix elements of the pairing force~(\ref{eq.9}) between Bloch states are given by an 
integral over the W-S cell of volume ${\cal V}_{\rm cell}$ 
\begin{equation}
\label{eq.10}
V_{\alpha \pmb{k}\beta \pmb{k^\prime}} =\int_{\rm WS} {\rm d}^3r\, v^{\pi}[\rho_n(\pmb{r})]\, 
|\varphi_{\alpha\pmb{k}}(\pmb{r})|^2 |\varphi_{\beta\pmb{k^\prime}}(\pmb{r})|^2\, ,
\end{equation}
the Bloch wavefunctions $\varphi_{\alpha\pmb{k}}(\pmb{r})$ being normalized according to
\begin{equation}
\label{eq.11}
\int_{\rm WS} {\rm d}^3r\, |\varphi_{\alpha\pmb{k}}(\pmb{r})|^2 = 1\, .
\end{equation}
The W-S cell that we consider here is a truncated octahedron, as determined by the body-centered 
cubic lattice geometry. It should not be confused with the spherical cell used in the 
W-S approximation~\cite{cha07}.

\begin{table}
\centering
\caption{Skyrme parameters of the force BSk16~\cite{cha08}. $\varepsilon_\Lambda$ 
is a s.p. energy cutoff above the chemical potential introduced in order to regularize 
the divergences associated with the zero range of the pairing force.} 
\label{tab1}
\vspace{.5cm}
\begin{tabular}{|c|c|}
\hline
  $t_0$ {\scriptsize [MeV fm$^3$]}   & -1837.23  \\
  $t_1$ {\scriptsize [MeV fm$^5$]}   & 383.521   \\
  $t_2$ {\scriptsize [MeV fm$^5$]}   & -3.41736 \\
  $t_3$ {\scriptsize [MeV fm$^{3+3\gamma}$]}  & 11523.0  \\
  $x_0$                              &  0.432600  \\
  $x_1$                              & -0.824106 \\
  $x_2$                              & 44.6520  \\
  $x_3$                              &  0.689797  \\
  $W_0$ {\scriptsize [MeV fm$^5$]}   &  141.100    \\
  $\gamma$                           &  0.3   \\
  $\varepsilon_\Lambda$	[MeV]             & 16 \\
\hline
\end{tabular}
\end{table}

\section{Neutron pairing gaps}
\label{sec4}
 
\begin{figure*}
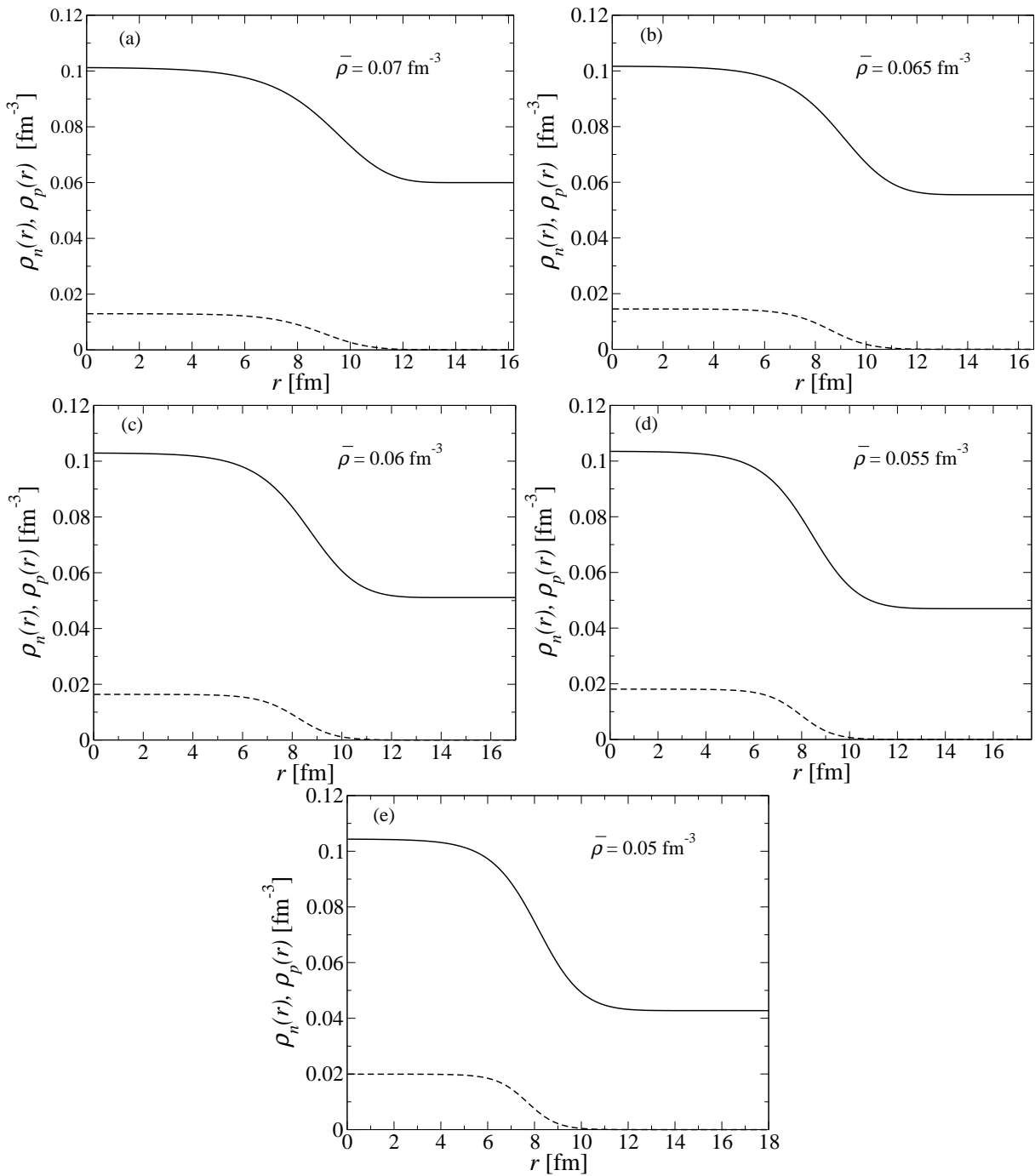

\begin{center}
\includegraphics[scale=0.3]{fig2a}
\includegraphics[scale=0.3]{fig2b}
\includegraphics[scale=0.3]{fig2c}
\includegraphics[scale=0.3]{fig2d}
\includegraphics[scale=0.3]{fig2e}
\end{center}
\vskip -0.5cm
\caption{Equilibrium neutron number density $\rho_n(r)$ (solid line) and proton number density $\rho_p(r)$ (dashed line) 
inside the Wigner-Seitz cell of different neutron-star crust layers with average nucleon density $\bar\rho$. 
The densities, expressed in fm$^{-3}$, have been obtained from the ETFSI method~\cite{onsi08} at $T=0$ 
with the Skyrme force BSk16~\cite{cha08}.
}
\label{fig2}
\end{figure*}

We have considered five different layers of the inner crust in the average nucleon 
density range between 0.05 fm$^{-3}$ and 0.07 fm$^{-3}$. Results of the ETFSI 
calculations at $T=0$ are summarized in Table~\ref{tab2} and the nucleon density profiles 
are plotted in Fig.~\ref{fig2}.

\begin{table}
\centering
\caption{Ground-state composition of the neutron-star crust using the ETFSI method 
with Skyrme force BSk16. $\bar\rho$ is the average nucleon density, $Z$ and $A$ the 
equilibrium numbers of protons and nucleons in the W-S cell respectively, and 
$\rho_{Bn}$ the neutron background density outside clusters (see Ref.~\cite{onsi08}).
}
\label{tab2}
\vspace{.5cm}
\begin{tabular}{|c|c|c|c|}
\hline
$\bar\rho$ [fm$^{-3}$] & $Z$ & $A$ & $\rho_{Bn}$ [fm$^{-3}$] \\
\hline
0.070 & 40 & 1258 & 0.060   \\
0.065 & 40 & 1264 & 0.056   \\
0.060 & 40 & 1260 & 0.051   \\ 
0.055 & 40 & 1294 & 0.047   \\ 
0.050 & 40 & 1304 & 0.043  \\ 
\hline
\end{tabular}
\end{table}

We have solved iteratively the anisotropic multi-band BCS gap equations~(\ref{eq.3}) 
for each average nucleon density $\bar\rho$. 
Results are summarized in Table~\ref{tab3}. 
Due to the ultra-violet divergence induced by the zero range of 
the pairing interaction~(\ref{eq.9}), the summation in Eq.~(\ref{eq.3}) has to be truncated. We have 
imposed the same s.p. energy cutoff $\varepsilon_\Lambda=16$ MeV above the chemical 
potential as used in the determination of the BSk16 force through optimization 
of the mass fit. For each temperature the chemical potential has been recalculated neglecting pairing 
since the pairing gaps are much smaller than the Fermi energy (note that the same approximation was used
in Ref.~\cite{cha08} to construct the effective density-dependent pairing strength from the pairing gap 
in infinite homogeneous neutron matter). We have solved Eq.~(\ref{eq.5}) by expanding the s.p. wavefunctions 
into plane-waves
\begin{equation}
\varphi_{\alpha\pmb{k}}(\pmb{r})=\exp({\rm i}\, \pmb{k}\cdot\pmb{r})\sum_{\pmb{G}} \widetilde{\varphi}_{\alpha\pmb{k}}(\pmb{G})\exp({\rm i}\, \pmb{G}\cdot\pmb{r})
\end{equation}
in which $\pmb{G}$ are reciprocal lattice vectors. Since by definition
\begin{equation}
\exp({\rm i}\, \pmb{G}\cdot\pmb{\ell})=1
\end{equation}
for any vectors $\pmb{G}$ and $\pmb{\ell}$, the Bloch boundary conditions~(\ref{eq.6}) 
are automatically satisfied. We have included all Fourier components with reciprocal lattice vectors 
$\pmb{G}$ such that $\vert\pmb{k}+\pmb{G}\vert < Q$.
$Q$ has been adjusted so that the s.p. energies are computed with an accuracy of a few keV. 
We have evaluated the summation in Eq.~(\ref{eq.3}) using the special-point method~\cite{hama92}. 
We have also applied this method to compute the pairing matrix elements~(\ref{eq.10}). On general 
grounds one may expect that $|\varphi_{\alpha\pmb{k}}(\pmb{r})|^2$, and thereby 
$V_{\alpha \pmb{k}\beta \pmb{k^\prime}}$ and $\Delta_{\alpha\pmb{k}}$, are weakly dependent on 
$\pmb{k}$, since bound states are vanishingly small outside clusters where Bloch boundary conditions 
are imposed while continuum states depend on $\pmb{k}$ essentially through only a phase factor 
$\exp({\rm i} \pmb{k}\cdot\pmb{r})$. Indeed we have found that the summation in Eq.~(\ref{eq.3}) 
converges quickly with the number of special $\pmb{k}$-points. An error below 1\% for the 
averaged pairing gap $\Delta_{\rm F}$ can be reached with 30 $\pmb{k}$-points. Keeping just one term, 
corresponding to the mean-value point~\cite{bal73}, yields a result with a few percent precision, as can be seen in 
Table~\ref{tab3}. This method has indeed proved to be surprisingly accurate in solid state physics to compute the 
electron density and dielectric matrix~\cite{bal78}.
The convergence of the real-space integrations in Eq.~(\ref{eq.10}) is slower because of the oscillating 
behavior of the wavefunctions. We have checked that the solutions of the BCS gap equations converge to a 
few keV accuracy using 110 special $\pmb{r}$-points. 

\begin{figure*}
\begin{center}
\includegraphics[scale=0.25]{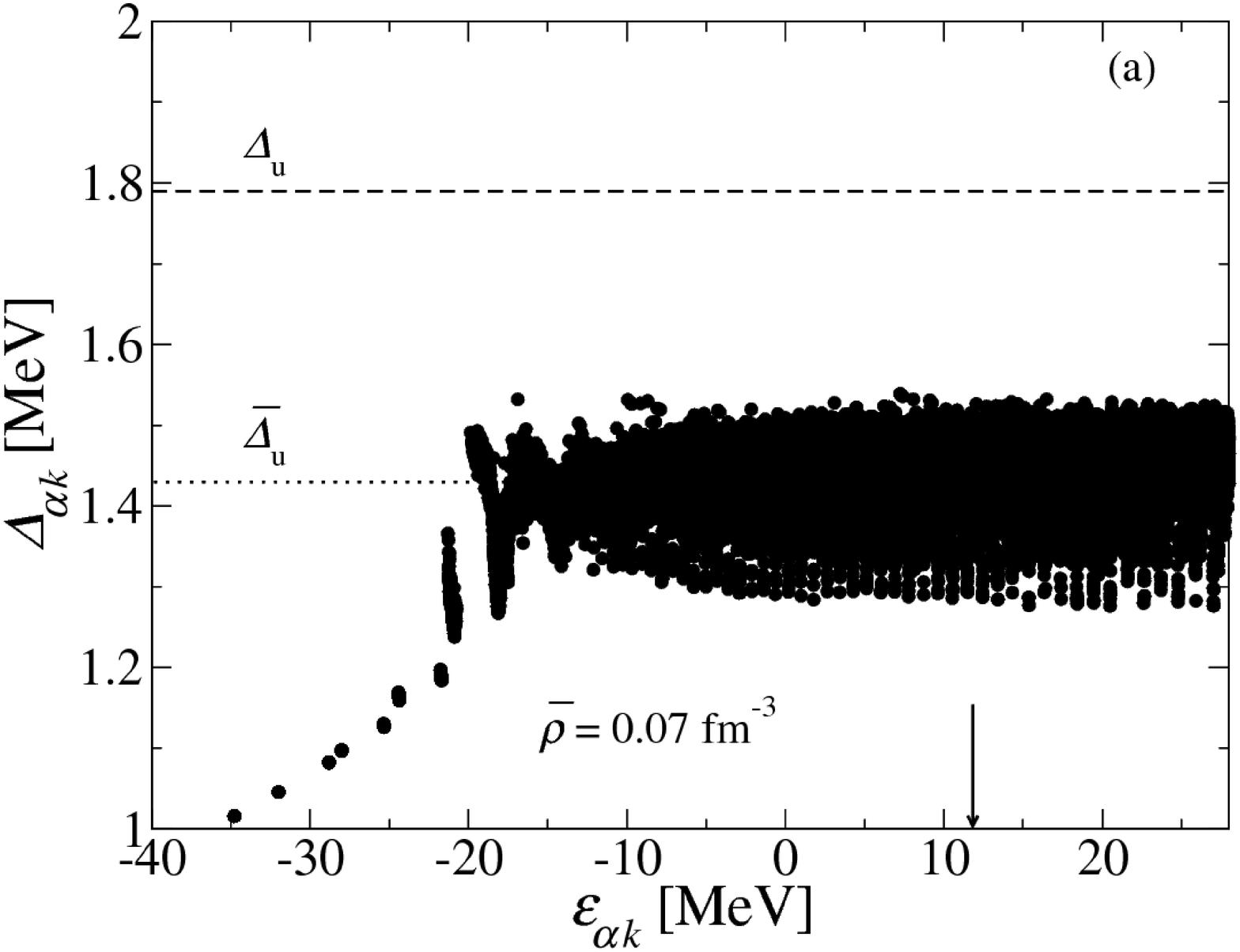}
\includegraphics[scale=0.25]{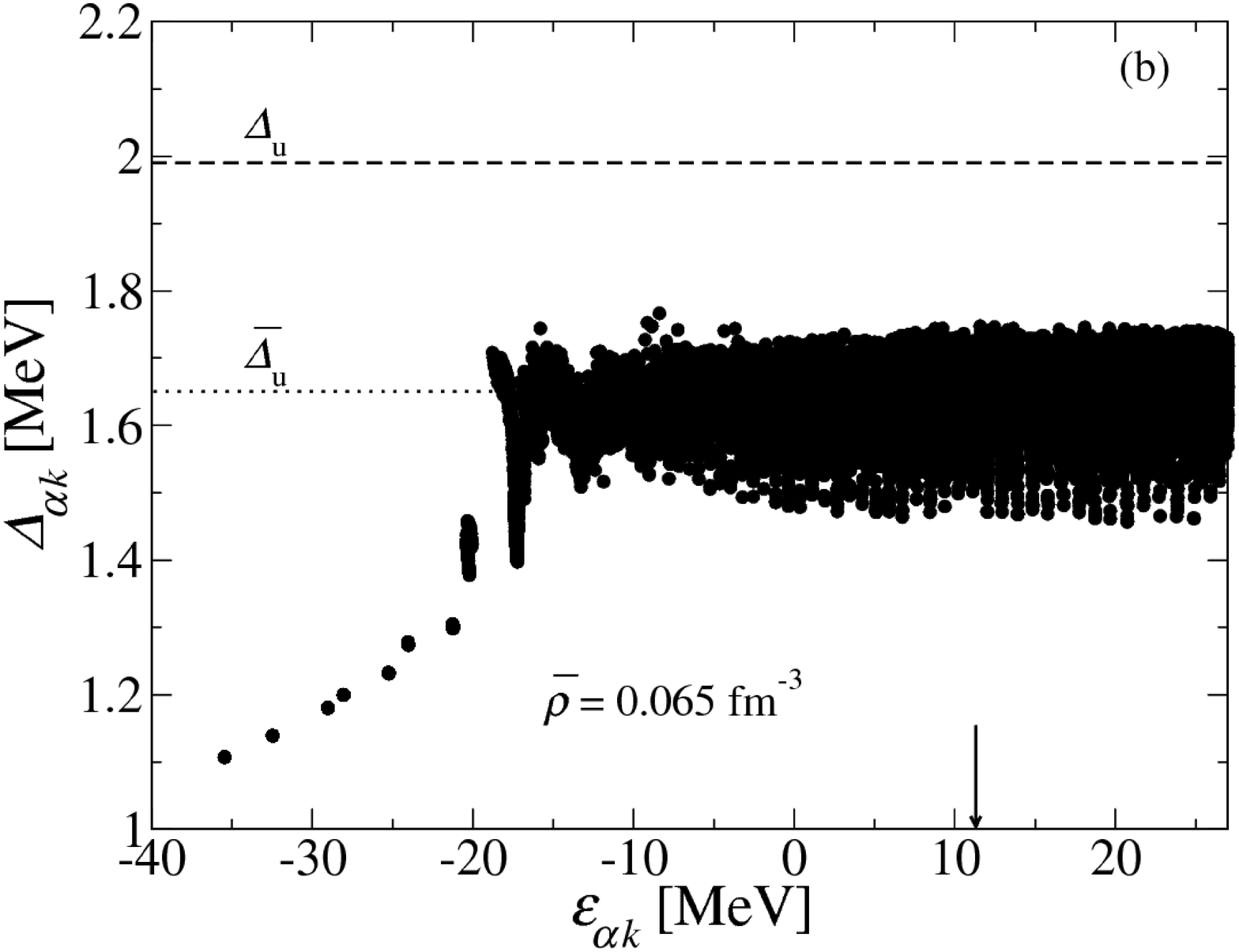}
\includegraphics[scale=0.25]{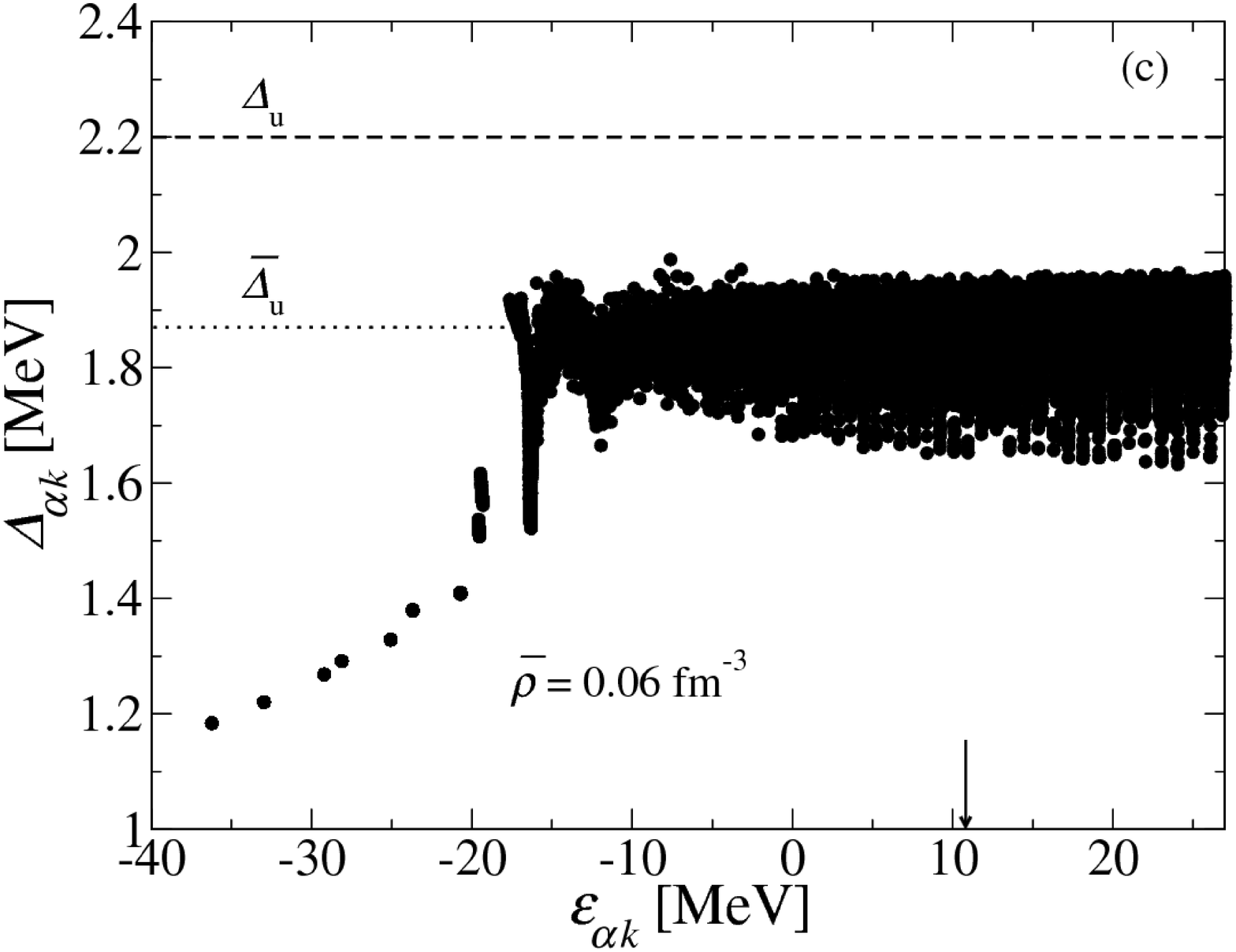}
\includegraphics[scale=0.25]{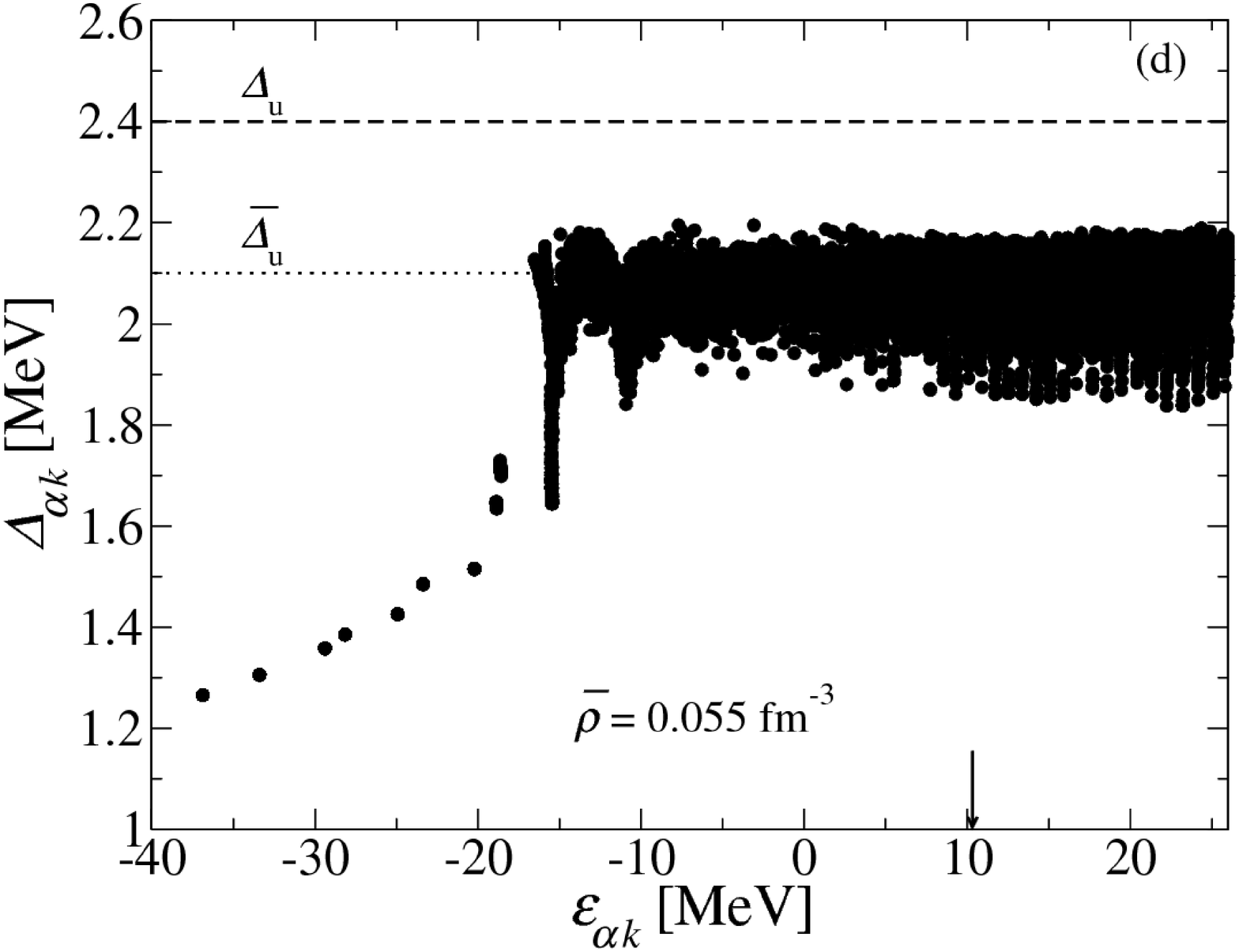}
\includegraphics[scale=0.25]{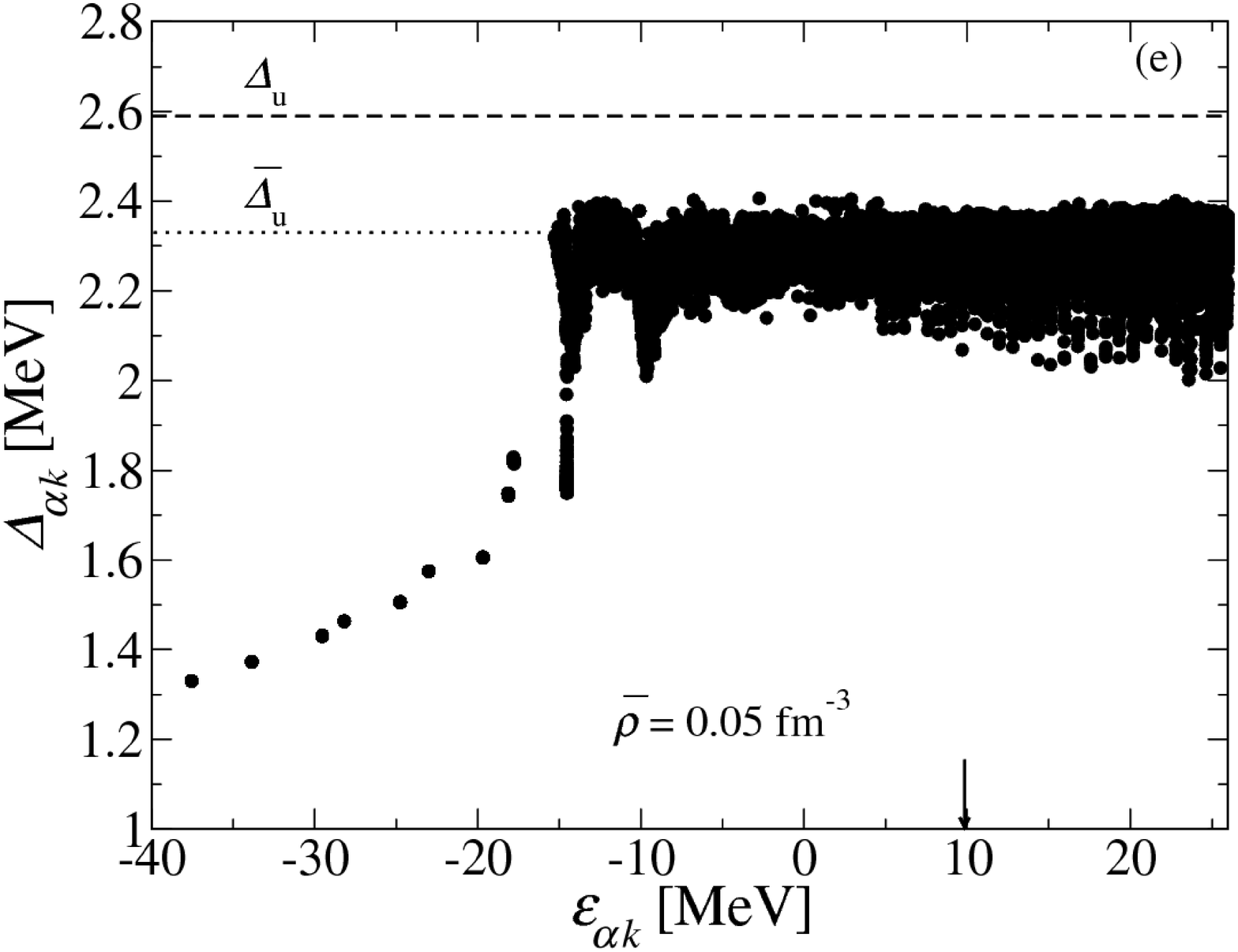}
\end{center}
\vskip -0.5cm
\caption{Neutron pairing gaps $\Delta_{\alpha\pmb{k}}$ vs s.p. energies 
$\varepsilon_{\alpha\pmb{k}}$ for the different crustal
layers. The arrow indicates the position of the chemical potential at $T=0$. 
The BCS equations~(\ref{eq.3}) have been solved at $T=0$ together 
with Eq.~(\ref{eq.5}) using the neutron-star crust composition shown in Fig.~\ref{fig2}. 
The dashed line and the dotted line represent the gaps $\Delta_{\rm u}$ and $\bar\Delta_{\rm u}$ 
respectively given in Table~\ref{tab3}.}
\label{fig3}
\end{figure*}

As can be seen in Fig.~\ref{fig3}, the dependence of the pairing gaps 
$\Delta_{\alpha\pmb{k}}$ on the band index $\alpha$ and wave vector $\pmb{k}$ is quite 
significant. At the Fermi level the pairing gaps vary by about $\sim 0.2-0.4$ MeV. 
While all s.p. states lying in the continuum contribute to the average gap 
$\Delta_{\rm F}$, we have found that $\Delta_{\rm F}$ remains almost unchanged if bound 
states are excluded from the summation in Eq.~(\ref{eq.3}). However this does not imply that  
neutrons inside clusters do not have any impact on the average pairing gap. The 
inhomogeneous distribution of neutrons modifies the s.p. energies and the matrix 
elements of the pairing force. 

For comparison, we have calculated the neutron pairing gap $\Delta_{\rm u}$ without nuclear 
clusters, assuming that unbound neutrons are uniformly distributed with the density $\rho_{Bn}$ 
corresponding to the neutron background density outside clusters (values of $\rho_{Bn}$ for
the different crustal layers are indicated in Table~\ref{tab2}). In this limiting case, 
the Bloch wavefunctions reduce to plane waves 
\begin{equation}
\label{eq.12}
\varphi_{\alpha\pmb{k}}(\pmb{r})=\frac{1}{{\cal V}_{\rm cell}}\exp[{\rm i}\, (\pmb{k}+\pmb{G_\alpha})\cdot\pmb{r}] \, , 
\end{equation}
where $\pmb{G_\alpha}$ are reciprocal lattice vectors. The pairing matrix elements become all 
equal
\begin{equation}
\label{eq.13}
V_{\alpha \pmb{k}\beta \pmb{k^\prime}} = \frac{v^{\pi}[\rho_{Bn}]}{{\cal V}_{\rm cell}} \, .
\end{equation}
As a result, the pairing gaps are independent of $\alpha$ and $\pmb{k}$, and are the solutions 
of the usual isotropic BCS equations~\cite{bcs57}
\begin{equation}
\label{eq.14}
1=-\frac{v^{\pi}[\rho_{Bn}]}{8\pi^2} \left(\frac{2 M_{Bn}^*}{\hbar^2}\right)^{3/2} 
\int_0^{\varepsilon_{B\rm F}+\varepsilon_\Lambda} {\rm d}\varepsilon 
\frac{\sqrt{\varepsilon}}{E(\varepsilon)}\tanh \frac{E(\varepsilon)}{2T} \, ,
\end{equation}
with 
\begin{equation}
E(\varepsilon)=\sqrt{(\varepsilon-\varepsilon_{B\rm F})^2+\Delta_{\rm u}^2}\, ,
\end{equation}
in which
\begin{equation}
\varepsilon_{B\rm F}=\frac{\hbar^2 k_{{\rm F} B}^2}{2 M_{Bn}^*}
\end{equation}
is the Fermi energy, $k_{{\rm F} B}=(3\pi^2 \rho_{Bn})^{1/3}$ and $M_{Bn}^*$ is the effective mass in neutron 
matter at density $\rho_{Bn}$. 
From the definition of the pairing strength, it follows immediately that $\Delta_u$ at $T=0$, 
is nothing else but the microscopic neutron pairing gap, shown in Fig.~\ref{fig1}, evaluated 
at the neutron density $\rho_n=\rho_{Bn}$. Results for the different crust layers are shown in
Table~\ref{tab3}. 
Comparing $\Delta_{\rm F}$ and $\Delta_{\rm u}$, it can be seen that the presence of inhomogeneities 
lowers the averaged neutron pairing gap by $\sim 10-20\%$. This reduction is much smaller than 
that found in previous calculations based on the W-S approach~\cite{bal07}. For instance, for 
$\bar\rho=0.058$ fm$^{-3}$, $\Delta_{\rm F}$ was suppressed by $\sim 40-60\%$ depending on the choice of 
boundary conditions (last line of Table 2 in Ref.~\cite{bal07}).

\begin{table}
\centering
\caption{Neutron pairing gaps in the neutron star crust for different average nucleon 
density $\bar\rho$ at $T=0$. $\Delta_{\rm F}$ ($\Delta_{\rm F0}$) is the pairing gap 
obtained from Eq.~(\ref{eq.3}) after averaging over continuum states and using 
the special point method (mean value point method) for the summation 
over $\pmb{k}$. $\Delta_{\rm u}$ and $\bar\Delta_{\rm u}$ are the pairing gaps in uniform 
infinite neutron matter for the neutron density $\rho_{Bn}$ and for the average neutron 
density $\bar\rho_n =(A-Z)/{\cal V}_{\rm cell}$ respectively.}
\label{tab3}
\vspace{.5cm}
\begin{tabular}{|c|c|c|c|c|c|c|}
\hline
$\bar\rho$ [fm$^{-3}$] &  $\Delta_{\rm F}$ [MeV] & $\Delta_{\rm F0}$ [MeV] & $\Delta_{\rm u}$ [MeV] & $\bar\Delta_{\rm u}$ [MeV]\\
\hline
0.070 &  1.44 & 1.39 & 1.79 & 1.43 \\ 
0.065 &  1.65 & 1.59 & 1.99 & 1.65 \\ 
0.060 &  1.86 & 1.81 & 2.20 & 1.87 \\ 
0.055 &  2.08 & 2.07 & 2.40 & 2.10 \\ 
0.050 &  2.29 & 2.27 & 2.59 & 2.33 \\ 
\hline
\end{tabular}
\end{table}

\section{Pairing field of the neutron superfluid at zero temperature and local density approximation}
\label{sec5}

\begin{figure*}
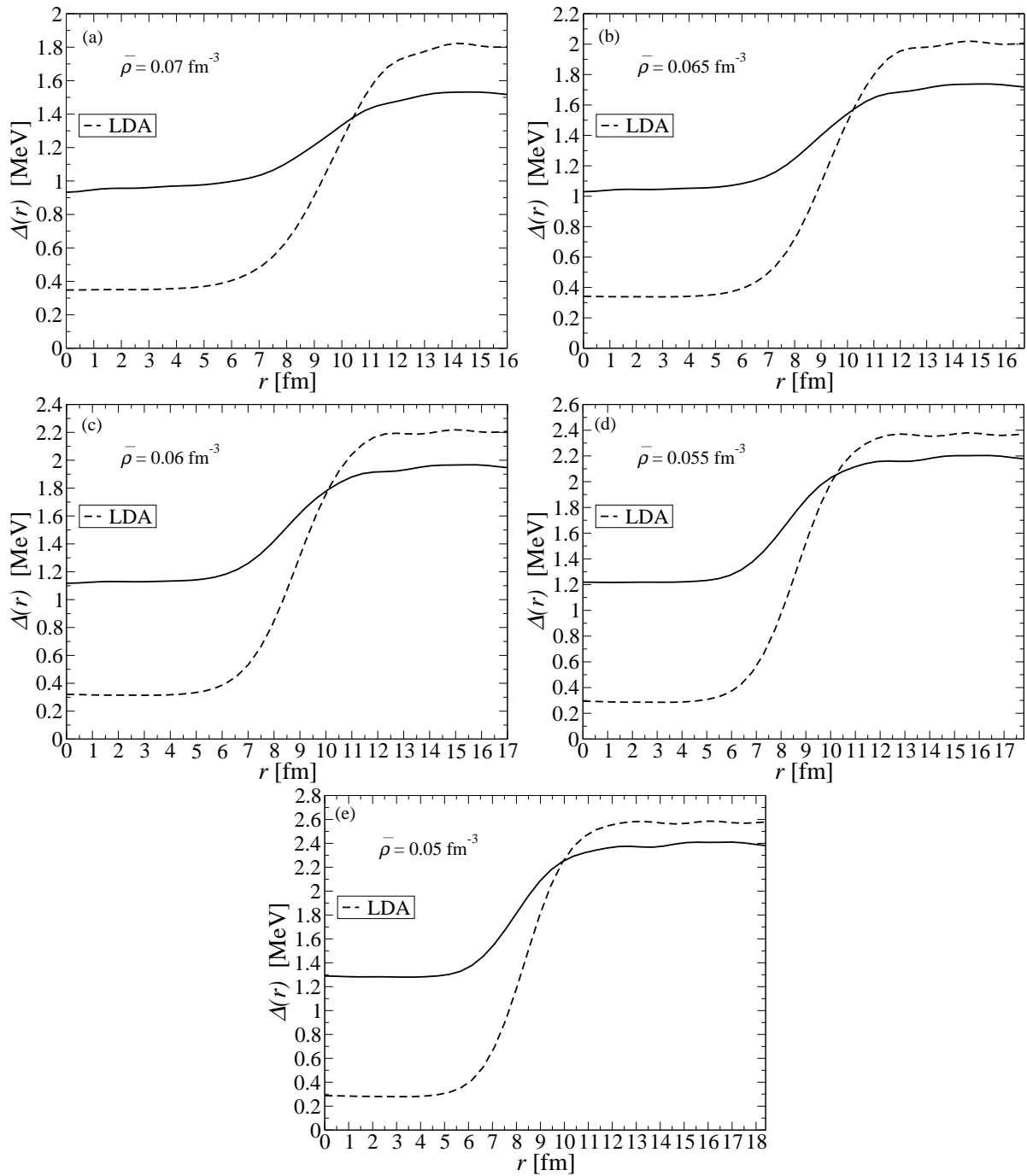

\begin{center}
\includegraphics[scale=0.3]{fig4a}
\includegraphics[scale=0.3]{fig4b}
\includegraphics[scale=0.3]{fig4c}
\includegraphics[scale=0.3]{fig4d}
\includegraphics[scale=0.3]{fig4e}
\end{center}
\vskip -0.5cm
\caption{Angle-averaged neutron pairing field $\Delta(r)$ (solid line) and LDA pairing field (dashed line) 
for the different crustal layers at $T=0$.}
\label{fig4}
\end{figure*}

\begin{figure*}
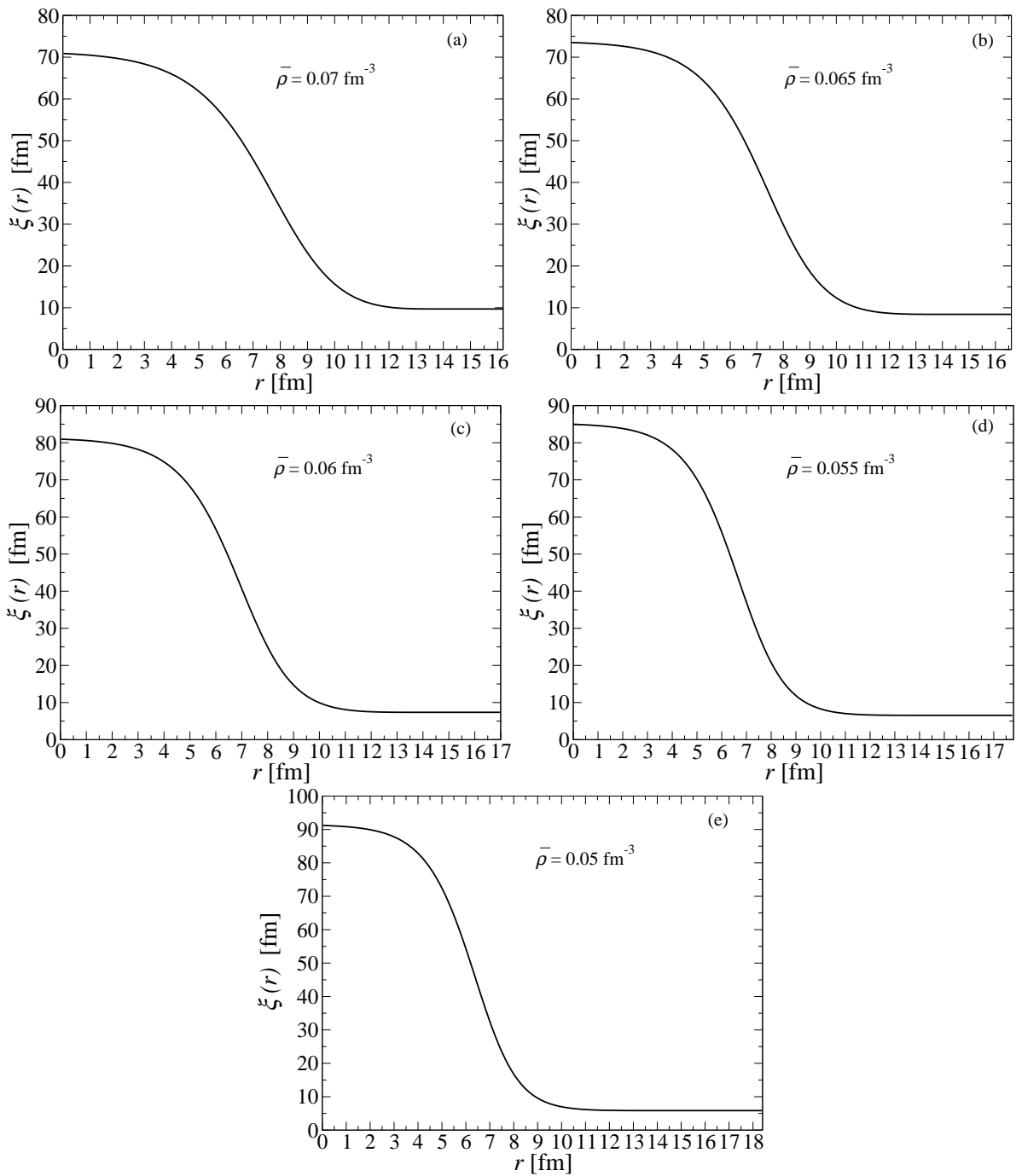

\begin{center}
\includegraphics[scale=0.3]{fig5a}
\includegraphics[scale=0.3]{fig5b}
\includegraphics[scale=0.3]{fig5c}
\includegraphics[scale=0.3]{fig5d}
\includegraphics[scale=0.3]{fig5e}
\end{center}
\vskip -0.5cm
\caption{Local coherence length, defined by Eq.~(\ref{eq.34}), 
for the different crustal layers at $T=0$.}
\label{fig5}
\end{figure*}

The effects of the inhomogeneities on the neutron superfluid can be more directly seen by computing the 
neutron pairing field, defined by 
\begin{equation}
\label{eq.29}
\Delta_n(\pmb{r})=-\frac{1}{2}v^{\pi n} [\rho_n(\pmb{r}),\rho_p(\pmb{r})]\tilde{\rho}_n(\pmb{r}) \, ,
\end{equation}
where $\rho_n(\pmb{r})$ and $\tilde{\rho}_n(\pmb{r})$ are the local normal neutron density and abnormal neutron density 
respectively given (at $T=0$) by
\begin{equation}
\label{eq.30}
\rho_n(\pmb{r})=\sum_{\alpha,\pmb{k}}^\Lambda |\varphi_{\alpha\pmb{k}}(\pmb{r})|^2 \biggl[1-\frac{\varepsilon_{\alpha\pmb{k}}-\mu}{E_{\alpha \pmb{k}}}\biggr]
\end{equation}
\begin{equation}
\label{eq.31}
\tilde{\rho}_n(\pmb{r})=\sum_{\alpha,\pmb{k}}^\Lambda |\varphi_{\alpha\pmb{k}}(\pmb{r})|^2 \frac{\Delta_{\alpha\pmb{k}}}{E_{\alpha \pmb{k}} }\, ,
\end{equation}
where the superscript $\Lambda$ is to indicate that the summation includes only states
whose s.p. energy lies below $\mu+\varepsilon_\Lambda$. 
We have computed the angle-averaged pairing field inside each W-S cell, as given by 
\begin{equation}
\label{eq.32}
\Delta_n(r)=\int_{\rm WS} \frac{{\rm d}\Omega}{4\pi}\, \Delta_n(\pmb{r})\, .
\end{equation}
For each value of the radial coordinate $r$, we have performed the solid-angle 
integration using 30 uniformly distributed spiral points on the sphere of radius $r$~\cite{rak94}.
In order to minimize the amount of computations, we have applied the mean-value point method for solving
the BCS equations and for calculating the normal and abnormal densities. For comparison, we have also 
calculated the pairing field in the LDA, i.e., assuming that at each point $\pmb{r}$ the pairing field 
is locally the same as that in uniform neutron matter for the density 
$\rho_n(\pmb{r})$
\begin{equation}
\label{eq.33}
\Delta^{({\rm LDA})}_n(\pmb{r})=\Delta_{\rm u}(\rho_n(\pmb{r}))\, .
\end{equation}
As can be seen in Fig.~\ref{fig4}, the LDA overestimates the spatial 
dependence of the pairing field thus indicating that pairing is highly non-local. 
Indeed the local coherence length of the neutron superfluid, defined by~\cite{bcs57}
\begin{equation}
\label{eq.34}
\xi(\pmb{r})=\frac{\hbar^2 k_{\rm F}(\pmb{r})}{\pi M_n^*(\pmb{r})\Delta^{({\rm LDA})}_n(\pmb{r})} \, ,
\end{equation}
where
\begin{equation}
\label{eq.34b}
k_{\rm F}(\pmb{r})=(3\pi^2\rho_n(\pmb{r}))^{1/3}\, ,
\end{equation}
is larger than the size of the clusters, as shown in Fig.~\ref{fig5}.
This means that even if the center of mass of 
a Cooper pair is located outside clusters, one of the partner may actually lie inside so that all neutrons 
are actually involved in the pairing process. As a result, pairing correlations are strongly enhanced 
inside clusters but are reduced in the intersticial region, leading to a smooth spatial variation of the 
pairing field. These so-called proximity effects are the most spectacular in the shallowest layer at 
$\bar\rho=0.05$ fm$^{-3}$ where the neutron pairing field is increased by a factor $\sim 4$ inside clusters. 
The entire inner crust is therefore permeated by the neutron superfluid, including in the region occupied by 
clusters themselves. 

The previous considerations suggest to estimate $\Delta_{\rm F}$ by calculating the 
pairing gap $\bar \Delta_{\rm u}$ in uniform neutron matter for the average 
neutron density $\bar\rho_n\equiv N/{\cal V}_{\rm cell}$ instead of the neutron background
density $\rho_{Bn}$. As can be seen in Table~\ref{tab3}, this gap indeed provides a 
very good approximation to the average gap calculated numerically.

\section{Critical temperature}
\label{sec6}

While the pairing gaps obtained from Eqs.~(\ref{eq.3}) vary from one band to the other as
shown in Fig.~\ref{fig3}, we have found that they all share the same universal temperature 
dependence which (for $^1$S$_0$ pairing) can be well-represented by~\cite{gor96}
\begin{equation}
\label{eq.15}
\Delta_{\alpha\pmb{k}}(T\leq T_{\rm c})\simeq \Delta_{\alpha\pmb{k}}(0) \sqrt{1-\left(\frac{T}{T_{\rm c}}\right)^{\delta}}\, ,
\end{equation}
with $\delta\simeq 3.23$ and the critical temperature $T_{\rm c}$, is defined by 
the condition $\Delta_{\alpha\pmb{k}}(T\geq T_{\rm c})=0$ for all states. 
For simplicity, we have assumed that the composition remains unchanged at finite temperature 
so that the s.p. energies $\varepsilon_{\alpha\pmb{k}}$ and the matrix elements $V_{\alpha \pmb{k}\beta \pmb{k^\prime}}$ 
are independent of $T$. We have determined 
$T_{\rm c}$ for each crustal layer by solving numerically Eqs.~(\ref{eq.3}). For comparison, 
we have also calculated the critical temperature $T_{\rm c u}$ of a uniform neutron superfluid 
with the corresponding neutron background density $\rho_{Bn}$. As indicated in Table~\ref{tab4}, 
the actual critical temperature $T_{\rm c}$ is systematically lower than $T_{\rm c u}$, as could have been 
expected from the results on the pairing gaps discussed in Section~\ref{sec5}. The critical temperature 
$\bar T_{\rm c u}$ of a uniform neutron superfluid with the average neutron density $\bar \rho_{n}$ is
much closer to $T_{\rm c}$. 

It is well-known from the BCS theory of superconductivity that the ratio of the critical temperature 
to the pairing gap is universal~\cite{bcs57}. We have checked that this relation holds in uniform 
neutron matter, namely
\begin{equation}
T_{\rm c u}=\frac{\exp(\zeta)}{\pi}\Delta_{\rm u}\, ,
\end{equation}
(likewise for $\bar T_{\rm cu}$) where $\zeta\simeq 0.577$ is the Euler-Mascheroni constant 
(this result is not specific to the pairing model
we have used. It has also been found in microscopic calculations using realistic nucleon-nucleon interactions,
as discussed in Ref.~\cite{lomb99}). What is perhaps more surprising is that we have found the same relation 
for the inhomogeneous neutron superfluid in the neutron-star crust
\begin{equation}
\label{eq.16}
\frac{T_{\rm c}}{\Delta_{\rm F}}\simeq \frac{\exp(\zeta)}{\pi}
\end{equation} 
(note that in the original BCS theory~\cite{bcs57}, the superconductor is supposed to be isotropic and 
uniform). Since the band theory includes the limiting case in which all neutrons are bound inside clusters, 
we expect Eq.~(\ref{eq.16}) to remain valid in finite nuclei. Quite interestingly, this conclusion seems to 
be supported by self-consistent mean-field calculations in tin isotopes~\cite{khan07} even though in this case 
the averaged gap $\Delta_{\rm F}$ has to be suitably defined.

We can understand Eq.~(\ref{eq.16}) by considering a simple pairing model. Since Eq.~(\ref{eq.16}) holds
in very different situations, this means that unlike the state-dependent pairing gaps $\Delta_{\alpha\pmb{k}}$, 
the critical temperature is rather insensitive to the precise nature of s.p. states (see also Ref.~\cite{yang05} for 
a mathematical discussion in the context of multi-band superconductivity). This suggests replacing the 
pairing matrix $V_{\alpha \pmb{k}\beta \pmb{k^\prime}}$ by some constant coupling $\langle V\rangle/{\cal V}_{\rm cell}$. 
In this case the pairing gaps $\Delta_{\alpha\pmb{k}}$ become all equal to the same value $\Delta$ and 
Eq.~(\ref{eq.3}) reduces to the finite-temperature isotropic BCS gap equations
\begin{equation}
\label{eq.17}
1 = - \frac{1}{2} \frac{\langle V\rangle}{{\cal V}_{\rm cell}} \int_0^{\mu+\varepsilon_\Lambda}{\rm d}\varepsilon 
\frac{g(\varepsilon)}{E(\varepsilon)} \tanh \frac{E(\varepsilon)}{2T}\, ,
\end{equation}
where $E(\varepsilon)=\sqrt{(\varepsilon-\mu)^2+\Delta^2}$ is the q.p. energy 
and $g(\varepsilon)$ is the density of s.p. states (for a given 
spin state) defined by
\begin{equation}
\label{eq.18}
g(\varepsilon)=\sum_{\alpha\,, \pmb{k}} \delta( \varepsilon_{\alpha\pmb{k}} - \varepsilon )\, .
\end{equation}

The critical temperature $T_{\rm c}$ is mainly determined by unbound neutron s.p. states (see 
the discussion in Section~\ref{sec5}). However, as shown in a previous work~\cite{cha09}, the 
density of unbound neutron s.p. states is essentially unaffected by the inhomogeneities on 
an energy scale larger than a few hundred keV. Since the pairing gaps are typically of the order of MeV, 
we can replace $g(\varepsilon)$ by its expression in uniform neutron matter with density 
$\bar \rho_n$
\begin{equation}
\label{eq.19} 
g(\varepsilon)\simeq \frac{{\cal V}_{\rm cell}}{4\pi^2} \left(\frac{2\bar M_n^*}{\hbar^2}\right)^{3/2}\sqrt{\varepsilon}\, ,
\end{equation}
where $\bar M_{n}^*$ is the corresponding neutron effective mass. 
At $T=0$ the neutron chemical potential $\mu_n$ is given by the neutron Fermi energy 
\begin{equation}
\label{eq.20}
\varepsilon_{\rm F}=\frac{\hbar^2 k_{\rm F}^2}{2 \bar M_{n}^*}
\end{equation}
with $k_{\rm F}=(3\pi^2\bar\rho_n)^{1/3}$. 
Since $T_{\rm c} \ll \varepsilon_{\rm F}$, we take $\mu(T_{\rm c})\simeq \varepsilon_{\rm F}$
remembering that the lowest order correction is only of order $(T_{\rm c}/\varepsilon_{\rm F})^2$. 
Since the most important contribution to the integral in Eq.~(\ref{eq.17}) comes from
s.p. states lying in the vicinity of the Fermi surface, we replace $g(\varepsilon)$
by $g(\varepsilon_{\rm F})$. For $T=T_{\rm c}$, $\Delta=0$ and Eq.~(\ref{eq.17}) thus becomes
\begin{equation}
\label{eq.21}
-\frac{4\pi^2\hbar^2}{\langle V\rangle \bar M_n^* k_{\rm F}} =  
\int_{-\varepsilon_F/2 T_{\rm c}}^{\varepsilon_\Lambda/2 T_{\rm c}}{\rm d}x\frac{\tanh |x|}{|x|}\, .
\end{equation}
After remarking that $T_{\rm c}\ll\varepsilon_\Lambda$, the integral in Eq.~(\ref{eq.21}) can be solved 
analytically using 
\begin{equation}
\label{eq.22}
\int_0^y {\rm d}x\frac{\tanh x}{x} \simeq \log\left(\frac{4y}{\pi}\right) + \zeta
\end{equation}
for $y\gg 1$.
Replacing Eq.~(\ref{eq.22}) in Eq.~(\ref{eq.21}), we find that the critical temperature 
is given by
\begin{equation}
\label{eq.23}
T_{\rm c}=\frac{2\exp(\zeta)}{\pi}\sqrt{\varepsilon_{\rm F}\varepsilon_\Lambda} \exp\left(\frac{8\pi^2\hbar^2}{k_{\rm F} \bar M_n^* \langle V \rangle }\right)\, .
\end{equation}

On the other hand, we assume that the average pairing gap $\Delta_{\rm F}$ can 
be obtained from the solution of the BCS gap Eq.~(\ref{eq.3}) at $T=0$ after substituting 
the pairing matrix $V_{\alpha \pmb{k}\beta \pmb{k^\prime}}$ by the same pairing constant 
$\langle V\rangle/{\cal V}_{\rm cell}$ that was introduced previously for calculating 
$T_{\rm c}$. Eq.~(\ref{eq.3}) thus reduces to Eq.~(\ref{eq.17}) which, at $T=0$, reads
\begin{equation}
\label{eq.24}
-\frac{4\pi^2\hbar^2}{\langle V\rangle \bar M_n^* k_{\rm F}} =  
\int_{-\varepsilon_{\rm F}/\Delta}^{\varepsilon_\Lambda/\Delta}{\rm d}x\, (1+x^2)^{-1/2}\, ,
\end{equation}
where we have taken the density of states out of the integral.
Solving Eq.~(\ref{eq.24}) for the pairing gap yields
\begin{equation}
\label{eq.25}
\Delta_{\rm F}=\Delta=2\sqrt{\varepsilon_{\rm F}\varepsilon_\Lambda} \exp\left(\frac{8\pi^2\hbar^2}{k_{\rm F} \bar M_n^* \langle V \rangle }\right)\, .
\end{equation}
Comparing Eqs.~(\ref{eq.23}) and (\ref{eq.25}) leads to Eq.~(\ref{eq.16}).

\begin{table}
\centering
\caption{Critical temperature for the onset of neutron superfluidity in the neutron-star crust
for different average nucleon density $\bar\rho$ using different approximations: $T_{\rm c}$ is the 
critical temperature obtained after solving numerically the BCS Eqs.~(\ref{eq.3}) while $T_{\rm c u}$ and 
$\bar T_{\rm c u}$ are the critical temperatures of a uniform superfluid with density $\rho_{Bn}$
and $\bar \rho_n$ respectively. All temperatures are indicated in units $10^{10}$ K.}
\label{tab4}
\vspace{.5cm}
\begin{tabular}{|c|c|c|c|c|c|c|}
\hline
$\bar\rho$ [fm$^{-3}$] & $T_{\rm c u}$ &  $T_{\rm c}$ &  $\bar T_{\rm c u}$ \\
\hline
0.070 & 1.17  & 0.96 &  0.94 \\ 
0.065 & 1.31  & 1.09 &  1.08 \\ 
0.060 & 1.45  & 1.24 &  1.23 \\ 
0.055 & 1.58  & 1.38 &  1.38 \\ 
0.050 & 1.70  & 1.52 &  1.53 \\ 
\hline
\end{tabular}
\end{table}

\section{Finite temperature effects on the neutron pairing field}
\label{sec7}

\begin{figure*}
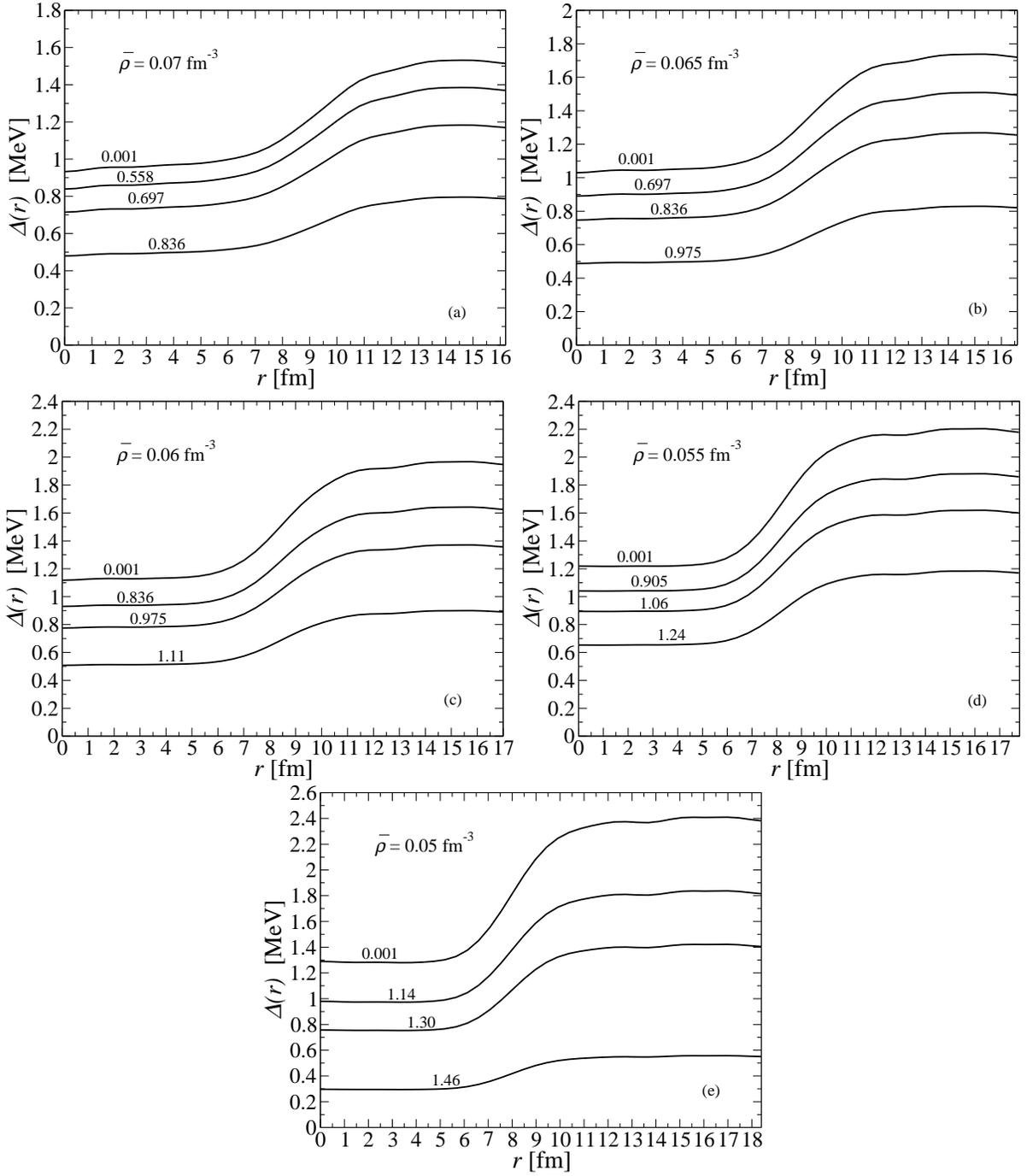

\begin{center}
\includegraphics[scale=0.3]{fig6a}
\includegraphics[scale=0.3]{fig6b}
\includegraphics[scale=0.3]{fig6c}
\includegraphics[scale=0.3]{fig6d}
\includegraphics[scale=0.3]{fig6e}
\end{center}
\vskip -0.5cm
\caption{Angle-averaged neutron pairing field $\Delta(r)$ for the different
crustal layers at different temperatures ($T$ is indicated above each curve in $10^{10}$ K).}
\label{fig6}
\end{figure*}

At finite temperatures, the pairing field is still given by the same expression~(\ref{eq.29}) as
for $T=0$. But the normal neutron density and abnormal neutron density are now given by 
\begin{equation}
\label{eq.30b}
\rho_n(\pmb{r})=\sum_{\alpha,\pmb{k}}^\Lambda |\varphi_{\alpha\pmb{k}}(\pmb{r})|^2 \biggl[1-\frac{\varepsilon_{\alpha\pmb{k}}-\mu}{E_{\alpha \pmb{k}}}\tanh \frac{E_{\alpha \pmb{k}}}{2T}\biggr]
\end{equation}
\begin{equation}
\label{eq.31b}
\tilde{\rho}_n(\pmb{r})=\sum_{\alpha,\pmb{k}}^\Lambda |\varphi_{\alpha\pmb{k}}(\pmb{r})|^2 \frac{\Delta_{\alpha\pmb{k}}}{E_{\alpha \pmb{k}} } \tanh \frac{E_{\alpha \pmb{k}}}{2T}\, .
\end{equation}
In the LDA, the corresponding pairing field is given by
\begin{equation}
\label{eq.33b}
\Delta^{({\rm LDA})}_n(\pmb{r},T)=\Delta_{\rm u}(\rho_n(\pmb{r}))\sqrt{1-\left(\frac{T}{T_{\rm cu}(\pmb{r})}\right)^\delta}
\end{equation}
for $T<T_{\rm cu}(\pmb{r})$ with
\begin{equation}
T_{\rm c u}(\pmb{r})=\frac{\exp(\zeta)}{\pi}\Delta_{\rm u}(\rho_n(\pmb{r}))\, ,
\end{equation}
while for $T>T_{\rm cu}(\pmb{r})$, $\Delta^{({\rm LDA})}_n(\pmb{r},T)=0$. 
Fig.~\ref{fig6} shows the pairing field for different temperatures. The temperature dependence 
of the pairing field comes mainly from that of the abnormal density. As the temperature gets closer 
to the critical temperature $T_{\rm c}$, the superfluid becomes more and more homogeneous. This is 
because the coherence length obtained after substituting $\Delta^{({\rm LDA})}_n(\pmb{r})$ by 
$\Delta^{({\rm LDA})}_n(\pmb{r},T)$ in Eq~(\ref{eq.34}), increases with $T$ and even diverges when the local 
critical temperature is reached. This means that the LDA becomes worse at finite temperatures, strongly 
overestimating the impact of inhomogeneities. In particular Eq~(\ref{eq.33b}) implies that 
$\Delta^{({\rm LDA})}_n(\pmb{r},T_{\rm cu}(0))$ vanishes inside clusters while it remains finite outside, 
leading to a sharp variation of the pairing field at the cluster surface. 

\section{Specific heat of superfluid neutrons}
\label{sec8}

Observations of the thermal X-ray emission from newly-born isolated neutron stars can 
potentially provide valuable information on the structure of neutron-star crusts. 
Due to its relatively low neutrino emissivity, the crust of the neutron star cools 
less rapidly than the core and thus stays hotter. As a result, the surface temperature 
decreases slowly during the first ten to hundred years after the formation of the neutron 
star in a supernova explosion and then drops sharply when the cooling wave from the core 
reaches the surface~\cite{lat94, gne01}. The occurrence of neutron superfluidity in the 
inner crust has a strong influence on the neutron specific heat and in turn, on the 
thermal relaxation time of the crust~\cite{monr07,for09}. 

\begin{figure*}
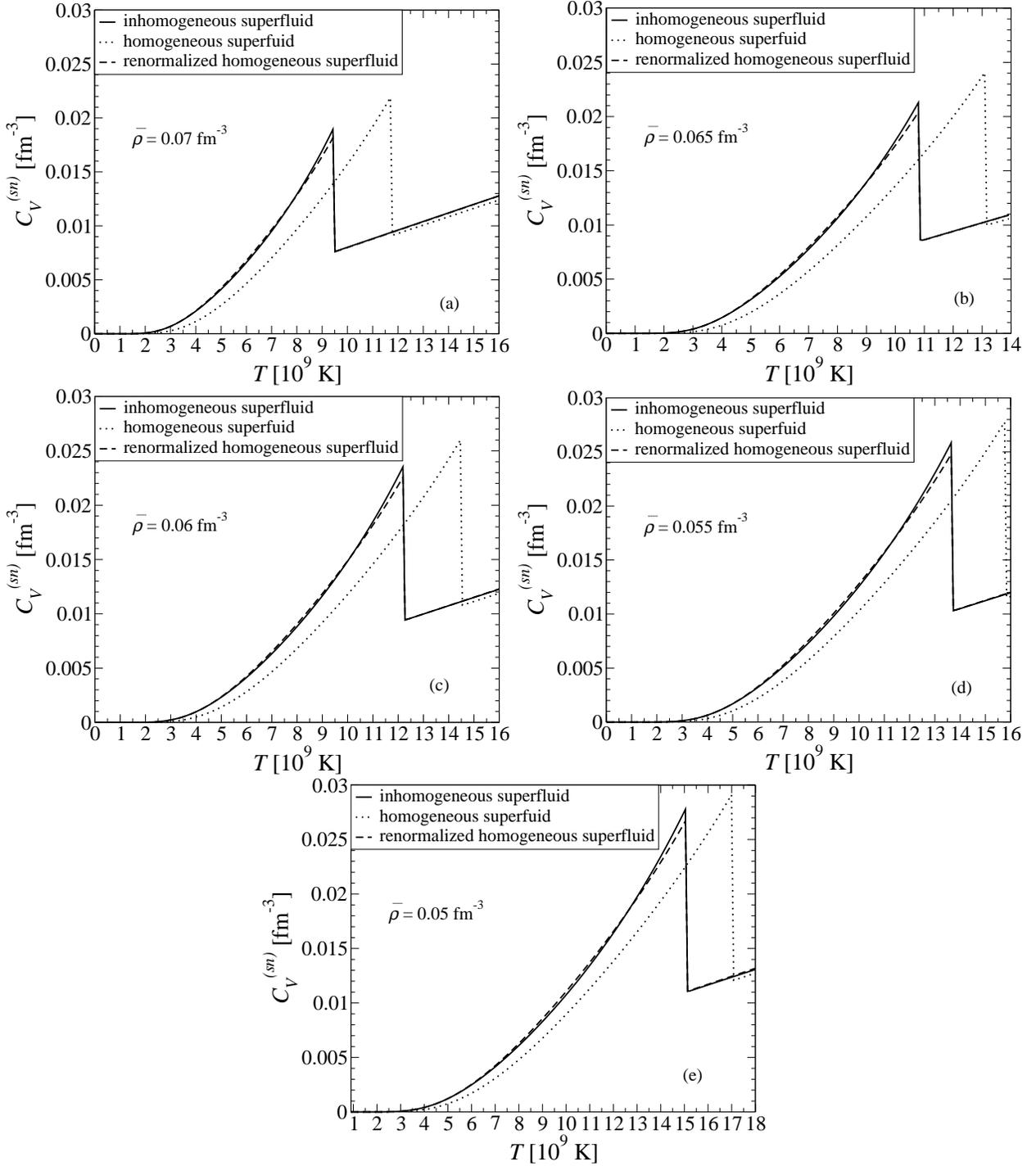

\begin{center}
\includegraphics[scale=0.3]{fig7a}
\includegraphics[scale=0.3]{fig7b}
\includegraphics[scale=0.3]{fig7c}
\includegraphics[scale=0.3]{fig7d}
\includegraphics[scale=0.3]{fig7e}
\end{center}
\vskip -0.5cm
\caption{Specific heat of superfluid neutrons in different layers of the inner crust 
of a neutron star. The solid line has been obtained from Eq.~(\ref{eq.39}) after solving
numerically the BCS Eqs.~(\ref{eq.3}) The dashed and dotted lines correspond to homogeneous 
neutron matter with two different values of the critical temperature $T_{\rm c}$ and $T_{\rm c u}$, 
respectively.}
\label{fig7}
\end{figure*}

In Ref.~\cite{cha09}, we have shown that the specific heat of normal neutrons in the outermost
layers of the inner crust is almost unaffected by the nuclear lattice and is essentially given
by the specific heat of a uniform gas. At low temperatures $T<T_{\rm c}$, pairing correlations 
have to be taken into account. This is usually done by introducing a multiplication factor $\cal R$, 
defined by 
\begin{equation}
\label{eq.36}
C_V^{(sn)}={\cal R}C_V^{(n)} \, ,
\end{equation}
where $C_V^{(sn)}$ and $C_V^{(n)}$ are the specific heat of superfluid and normal 
neutrons respectively. Assuming that the free neutrons are uniformly distributed, 
the factor $\cal R$ is well-approximated by the analytical expression of Ref.~\cite{lev94} 
(used in neutron-star cooling simulations)
\begin{eqnarray}
\label{eq.37}
{\cal R}(y)=\biggl[0.4186+\sqrt{1.007^2+(0.501 u)^2}\biggr] \nonumber\\
\times\exp(1.456-\sqrt{1.456^2+u^2})
\end{eqnarray}
with $y=T/T_{\rm cu}$ and 
\begin{equation}
\label{eq.38}
u=\sqrt{1-y}\biggl(1.456-\frac{0.157}{\sqrt{y}}+\frac{1.764}{y}\biggr)\, .
\end{equation}
In order to assess the validity of these expressions for the inhomogeneous matter of 
neutron-star crusts, we have evaluated the neutron specific heat from the numerical solutions of 
the anisotropic multi-band BCS gap equations~(\ref{eq.3}). We have applied the following expression 
from Ref.~\cite{degen66} using the temperature dependence of the gaps given by Eq.~(\ref{eq.15})
\begin{multline}
\label{eq.39}
C_V^{(sn)}=\frac{1}{{\cal V}_{\rm cell}}\sum_{\alpha,\pmb{k}}\frac{\exp{(E_{\alpha\pmb{k}}/T)}}{\left[1+\exp{(E_{\alpha\pmb{k}}/T)}\right]^2} \biggl[ \left(\frac{E_{\alpha\pmb{k}}}{T}\right)^2 \\
-\frac{1}{2T} \frac{{\rm d}}{{\rm d}T} \Delta_{\alpha\pmb{k}}(T)^2 \biggr]\, .
\end{multline}
Numerical results are shown in Fig.~\ref{fig7}. The summation in $\pmb{k}$-space in Eq.~(\ref{eq.39})
has been carried out using the special-point method~\cite{hama92}. 

At low temperatures $T\ll T_{\rm c}$, neutron pairing correlations are very strong. 
The pairing gaps are weakly dependent on $T$ and the last term of Eq.~(\ref{eq.39}) is 
negligible (de Genne approximation~\cite{degen66}). In this regime, the neutron specific heat is 
exponentially suppressed compared to that of normal neutrons and decays approximately 
like $\exp(-\Delta_{\rm F}/T)$. However as the temperature gets higher, the contribution of the last 
term of Eq.~(\ref{eq.39}) becomes increasingly large, eventually leading to a sharp rise of the 
specific heat for $T\lesssim T_{\rm c}$. For $T>T_{\rm c}$, all pairing gaps vanish and neutrons 
superfluidity is destroyed. For $T>T_{\rm c}$ and $T\ll \varepsilon_{\rm F}$, the specific heat increases 
linearly with $T$ as can be clearly seen in Fig.~\ref{fig7}. As in our previous work~\cite{cha09}, we have 
found that the numerical results for the normal neutron specific heat are in good agreement 
with the expression for a uniform gas
\begin{equation}
\label{eq.35}
C_V^{(n)}=\frac{\bar M_n^* k_{\rm F}}{3\hbar^2} T\, .
\end{equation}
Note however that for the dense crustal layers considered here, using 
$\bar M_n^*$ and $k_{\rm F}$ instead of $M_{Bn}^*$ and $k_{{\rm F} B}$ (as in 
Ref.~\cite{cha09}) yields a slightly better fit. 
As shown in Fig.~\ref{fig7}, the transition between the superfluid and the normal 
regime is very sharp. The specific heat exhibits a discontinuity at $T=T_{\rm c}$, 
approximately given by 
\begin{equation}
\label{eq.40}
\frac{C_V^{(sn)}(T_{\rm c})-C_V^{(n)}(T_{\rm c})}{C_V^{(n)}(T_{\rm c})}\simeq\frac{3}{2}\delta\exp(-2\zeta)\, ,
\end{equation}
after replacing $\Delta_{\alpha\pmb{k}}$ by $\Delta_{\rm F}$ in Eq.~(\ref{eq.39}) and 
the density of s.p. state by Eq.~(\ref{eq.19}). Note that for $T\lesssim T_{\rm c}$, 
the neutron specific heat is enhanced by pairing, i.e. ${\cal R}>1$. 

We have found that $C_V^{(sn)}$ is numerically close to the expression for a 
uniform superfluid, namely Eqs.~(\ref{eq.36}), (\ref{eq.37}) and (\ref{eq.35}), 
\emph{provided} that the critical temperature is suitably renormalized 
(i.e., $T_{\rm cu}$ must be replaced by $T_{\rm c}$) as shown in Fig.~\ref{fig7}
(renormalized homogeneous superfluid curve). 

\section{Conclusions}
 
We have clarified the effects of the nuclear inhomogeneities on the neutron superfluidity in the deep 
layers of the neutron-star crust by solving the anisotropic multi-band BCS gap equations~(\ref{eq.3}). 
We have properly taken into account the interactions between the neutron superfluid and the nuclear crystal
lattice by imposing Bloch boundary conditions instead of using the W-S approximation. 
Due to the presence of the nuclear clusters, neutrons belonging to different bands and having 
different Bloch wave vectors feel different pairing interactions thus leading to a dispersion of the neutron 
pairing gaps $\Delta_{\alpha\pmb{k}}$ of about $\sim 0.2-0.4$ MeV around the Fermi level, as shown in Fig.~\ref{fig3}. 
The neutron pairing gap $\Delta_{\rm F}$ \emph{averaged} over all continuum states is 
reduced due to the presence of inhomogeneities, but much less than predicted by previous calculations based on the W-S 
approximation~\cite{bal07}. Unlike the individual gaps $\Delta_{\alpha\pmb{k}}$, $\Delta_{\rm F}$ is 
essentially unaffected by the band structure and is very well approximated by the gap $\bar \Delta_{\rm u}$ in 
uniform neutron matter using the average neutron density $\bar \rho_n$. The reason for this agreement lies in 
the highly non-local character of the pairing phenomenon involving both bound and unbound neutrons, as revealed by 
numerical calculations of the neutron pairing field $\Delta_n(r)$. As a consequence the LDA strongly 
overestimates the spatial variations of the pairing field. The discrepancies are particularly large inside 
clusters where the LDA incorrectly predicts a quenching of pairing correlations. 

Solving the BCS equations at finite temperatures, we have found that the temperature dependence 
of the pairing gaps $\Delta_{\alpha\pmb{k}}$ is universal and well-represented by Eq.~(\ref{eq.15}). 
Unlike what one might have naively expected, the critical temperature $T_{\rm c}$ is not determined by 
the largest state-dependent pairing gap around the Fermi level, but rather by the average pairing 
gap $\Delta_{\rm F}$. Moreover the ratio $T_{\rm c}/\Delta_{\rm F}$ 
is approximately given by the BCS value $\simeq 0.58$~\cite{bcs57}. We have explained these results by 
solving the isotropic BCS Eqs.~(\ref{eq.17}) in the weak-coupling approximation leading to the 
analytical expression~(\ref{eq.23}) for the critical temperature. We have computed the neutron pairing 
field at finite temperatures and we have shown that the LDA becomes worse as the temperature 
is increased.

We have computed the neutron specific heat, which is an important ingredient for modeling the 
thermal evolution of newly-born neutron stars~\cite{lat94, gne01,monr07,for09}. We have found that
the specific heat is modified by band effects but it can be easily estimated from the expression in 
uniform neutron matter, namely Eqs.~(\ref{eq.36}), (\ref{eq.37}) and (\ref{eq.35}), 
by simply renormalising $T_{\rm c}$. 

The conclusions of the present work may change in shallower regions of the crust where the matter is more 
inhomogeneous, as suggested by a recent study in dirty superconductors~\cite{zou08}. In addition we have 
neglected the change in composition with increasing temperature~\cite{onsi08} which could affect the 
values of the pairing gaps (hence also the critical temperature) for $T\gtrsim 10^{10}$ K. In particular
we expect to find deviations of the ratio $T_{\rm c}/\Delta_{\rm F}$ from the BCS value$\simeq 0.58$ when 
these thermal effects are taken into account. We have also left aside the modifications of the pairing gaps 
due to many-body effects beyond the BCS approximation~\cite{vig05}. These various issues will be addressed 
in future works using the multi-band approach presented in this paper.

\begin{acknowledgments}
N.~C. acknowledges financial support from a Marie Curie Intra-European grant 
(contract number MEIF-CT-2005-024660). This work was also supported by FNRS (Belgium), 
NSERC (Canada) and by CompStar, a Research Networking Programme of the European Science 
Foundation.
\end{acknowledgments}


\begin{thebibliography}{99}
\bibitem{mig59} A. B. Migdal, Nucl. Phys. {\bf 13}, 655 (1959).
\bibitem{bcs57} J. Bardeen, L. N. Cooper, and J. R. Schrieffer, Phys. Rev. {\bf 108}, 1175 (1957).
\bibitem{baym69} G. Baym, C. J. Pethick, and D. Pines, Nature {\bf 224}, 673 (1969).
\bibitem{and75} P. W. Anderson and N. Itoh, Nature {\bf 256}, 25 (1975).
\bibitem{chac06} N. Chamel and B. Carter, Mon.Not.Roy.Astron.Soc. {\bf 368}, 796 (2006).
\bibitem{kos09} K. Glampedakis and N. Andersson, Phys. Rev. Lett. {\bf 102}, 141101 (2009).
\bibitem{lat94} J. M. Lattimer, K. A. van Riper, M. Prakash and M. Prakash, Astrophys. J.{\bf 425} 802 (1994).
\bibitem{gne01} O.Y. Gnedin, D.G. Yakovlev, A.Y. Potekhin, Mon.Not.Roy.Astron.Soc. {\bf 324}, 725 (2001).
\bibitem{monr07} C. Monrozeau, J. Margueron, N. Sandulescu, Phys. Rev. C {\bf 75} (2007) 065807.
\bibitem{for09} M. Fortin, F. Grill , J. Margueron, N. Sandulescu, arxiv preprint 0910.5488.
\bibitem{agui09} D. N. Aguilera, V. Cirigliano, J.A. Pons, S. Reddy, R. Sharma, Phys. Rev.Lett.{\bf 102}, 091101 (2009).
\bibitem{shter07} P. S. Shternin, D. G. Yakovlev, P. Haensel, A. Y. Potekhin, Mon. Not. Roy. Astr. Soc. Lett. {\bf 382}, L43 (2007).
\bibitem{brown09} E. F. Brown, A. Cumming, Astrophys. J. {\bf 698}, 1020 (2009).
\bibitem{car05} B. Carter, N. Chamel, P. Haensel, Nucl. Phys. {\bf A 748},675 (2005).
\bibitem{cha05} N. Chamel, Nucl.Phys. {\bf A747}, 109 (2005).
\bibitem{lars09} L. Samuelsson, N. Andersson, Class. Quant. Grav. {\bf 26}, 155016 (2009).
\bibitem{dean03} D. J. Dean and M. Hjorth-Jensen, Rev. Mod. Phys. {\bf 75}, 607 (2003).
\bibitem{blas97} F. V. de Blasio, M. Hjorth-Jensen, {\O}. Elgar{\o}y, L. Engvik,
G. Lazzari, M. Baldo, and H.-J. Schulze, Phys. Rev. C {\bf 56}, 2332 (1997).
\bibitem{mat06} M. Matsuo, Phys. Rev. C {\bf 73}, 044309 (2006).
 \bibitem{lrr} N. Chamel and P. Haensel,``Physics of Neutron Star Crusts'', Living Rev. Relativity 11, (2008), 10. 
http://www.livingreviews.org/lrr-2008-10
\bibitem{tin96} M. Tinkham, Introduction to Superconductivity, Mc Graw Hill (1996).
\bibitem{bar97} F. Barranco, R. A. Broglia, H. Esbensen, and E. Vigezzi,
Phys. Lett. {\bf B 390}, 13 (1997).
\bibitem{bar98} F. Barranco, R. A. Broglia, H. Esbensen, and E. Vigezzi,
Phys. Rev. C {\bf 58}, 1257 (1998).
\bibitem{mont04} F. Montani, C. May, and H. M\"uther, Phys. Rev. C {\bf 69}, 065801 (2004).
\bibitem{sand04} N. Sandulescu, N. V. Giai, and R. J. Liotta, Phys. Rev. C {\bf 69}, 045802 (2004) ; 
N. Sandulescu, Phys. Rev. C {\bf 70}, 025801 (2004).
\bibitem{khan05} E. Khan, N. Sandulescu, and N. V. Giai, Phys. Rev. C {\bf 71}, 042801(R) (2005). 
\bibitem{bal07} M. Baldo, E.E. Saperstein, S.V. Tolokonnikov, Eur.Phys.J. {\bf A 32}, 97 (2007).
\bibitem{bal06} M. Baldo, E.E. Saperstein, S.V. Tolokonnikov, Nucl. Phys. {\bf A 775}, 235 (2006).
\bibitem{cha07} N. Chamel, S. Naimi, E. Khan, J. Margueron, Phys. Rev. C{\bf 75} (2007) 055806. 
\bibitem{suhl59} H. Suhl, B. T. Matthias, and L. R. Walker, Phys. Rev. Lett. {\bf 3}, 552 (1959).
\bibitem{choi02} H. J. Choi, D. Roundy, H. Sun, M. L. Cohen, and S. G.
Louie, Nature {\bf 418}, 758 (2002).
\bibitem{iav02} M. Iavarone, G. Karapetrov, A. E. Koshelev, W. K.
Kwok, G. W. Crabtree, D. G. Hinks, W. N. Kang, E.-M. Choi, H. J. Kim, H.-J. Kim, et al., Phys. Rev. Lett. {\bf 89}, 187002 (2002).
\bibitem{hun08} F. Hunte, J. Jaroszynski, A. Gurevich, D. C. Larbalestier, R. Jin, A. S. Sefat, M. A. McGuire, B. C. Sales, D. K. Christen and 
D. Mandrus, Nature {\bf 453}, 903 (2008).
\bibitem{onsi08} M. Onsi, A. K. Dutta, H. Chatri, S. Goriely, N. Chamel, J. M. Pearson, Phys. Rev. C {\bf 77}, 065805 (2008).
\bibitem{oy94} K. Oyamatsu and M. Yamada, Nucl. Phys. {\bf A578}, 181 (1994).
\bibitem{cha08} N. Chamel, S. Goriely, and J. M. Pearson,   Nucl. Phys. {\bf A812}, 72 (2008).
\bibitem{hama92} J. Hama and M. Watanabe, J. Phys. Cond. Mat. {\bf 4}, 4583 (1992).
\bibitem{bal73} A. Baldereschi, Phys. Rev. B {\bf 7}, 5212 (1973).
\bibitem{bal78} A. Baldereschi and E. Tosatti, Phys. Rev. B {\bf 17}, 4710 (1978).
\bibitem{rak94} E. A. Rakhmanov, E. B. Saff and Y.M. Zhou, Math. Res. Lett. {\bf 1}, 647 (1994).
\bibitem{gor96} S. Goriely, Nucl. Phys. {\bf A605}, 28 (1996).
\bibitem{lomb99} U. Lombardo, Superfluidity in Nuclear Matter, in \textit{Nuclear Methods and  Nuclear
Equation of State}, ed. by M. Baldo (World Scientific, Singapore, 1999), 458-510.
\bibitem{khan07} E. Khan, Nguyen Van Giai, N. Sandulescu, Nucl. Phys. {\bf A 789} (2007), 94.
\bibitem{yang05} Y. Yang, Physica D {\bf 200}, 60 (2005).
\bibitem{cha09} N. Chamel, J. Margueron, E. Khan, Phys. Rev. C {\bf 79}, 012801(R) (2009).
\bibitem{lev94} K.~P. Levenfish and D.~G. Yakovlev, Astron. Rep. {\bf 38} (1994), 247.
\bibitem{degen66} P.~G. de Gennes, Superconductivity of Metals and Alloys, W.A.
Benjamin inc., New York Amsterdam (1966), p129.
\bibitem{zou08} Y. Zou, I. Klich and G. Refael, Phys. Rev. B {\bf 77} (2008), 144523.
\bibitem{vig05} E. Vigezzi, F. Barranco, R. A. Broglia, G. Col`o, G. Gori, and F. Ramponi, 
Nucl. Phys. {\bf A 752}, 600 (2005).


\end{thebibliography}
\end{document}